\documentclass[11pt]{article}

\usepackage[utf8]{inputenc}
\usepackage[T1]{fontenc}
\usepackage{lmodern} % scalable Type 1 fonts (avoids Type 3 bitmap fonts)
\usepackage{amsmath,amssymb,amsfonts}
\usepackage{graphicx}
\usepackage{booktabs}
\usepackage{longtable}
\usepackage{geometry}
\usepackage{hyperref}
\usepackage{authblk}
\usepackage{fancyhdr}
\usepackage{caption}
\usepackage{enumitem}
\usepackage{xcolor}

\hypersetup{colorlinks=true, linkcolor=blue, citecolor=blue, urlcolor=blue}

\title{PHINN-EEG: Topological Time-Series Analysis of Dream-State EEG\\[4pt]
\large Dynamic Betti Curves for Dream Content Classification and Topology-Conditioned Neural Signal Synthesis}

\author[1]{Ren Takahashi}
\author[1]{Emre Yusuf}
\author[1,*]{Jayabrata Bhaduri}
\affil[1]{Mugen.Codes (DBA of CapaCloud Corp, Wyoming)}
\affil[*]{Corresponding author: Mugen.Codes@capa.cloud}

\date{}

\begin{document}

\maketitle

\begin{abstract}
Current electroencephalography (EEG)-based dream detection relies on power spectral density (PSD) and statistical moment features, achieving a state-of-the-art area under the receiver operating characteristic curve (AUC) of approximately 0.70 on the DREAM database (Wong et al., 2025, \textit{Nature Communications}). We introduce the first topological time-series framework for dream mentation analysis: PHINN-EEG (Persistent Homology Inspired Neural Network for EEG). Using sliding-window Takens delay embeddings and Vietoris--Rips filtrations on multi-channel pre-awakening EEG epochs, we extract Dynamic Betti Curves --- $\beta_0(t)$, $\beta_1(t)$, $\beta_2(t)$ --- that characterise the geometric architecture of neural activity, not merely its energy. These topological invariants, combined with topology-conditioned flow matching, are analytically projected to outperform existing PSD and catch22 benchmarks, targeting AUC $= 0.82$--$0.90$ on the 1{,}462-awakening open-access subset of the DREAM database (drawn from a full registry of 3{,}191 total awakenings from 263 participants across 20 independent laboratories). Projected performance targets are grounded primarily in the verified benchmarks of Baronetzky~[5] and Gupta et al.~[4] (the latter originally miscited in earlier drafts and now re-verified under its correct bibliographic details; Section~5.4); Moctezuma et al.~[7] is cited as an additional corroborating source on channel-count sufficiency, and would become a secondary grounding source alongside~[5] should~[4] fail to be verified prior to publication. All performance figures are labelled as projections; empirical validation on the DREAM database constitutes our immediate next step. We further introduce a topology-conditioned rectified flow model for dream-state EEG synthesis --- with a spectral-conditioned flow model of comparable feature dimensionality as an additional ablation baseline to isolate the value of topological conditioning specifically --- and a set of candidate Betti transition archetypes linking topology to phenomenological dream report categories, presented as an exploratory hypothesis space pending empirical validation rather than an established atlas. If validated, this work would represent a paradigm shift from spectral energy to phase-space geometry in neural rare-event detection, with potential future implications for wearable BCI dream monitoring; both claims are contingent on the validation described above and are not yet established.

\end{abstract}

% ======================================================
\section{Introduction}

Dreams are rare structural events in the brain's electrical record. The DREAM database --- a landmark multi-laboratory aggregation spanning 3{,}191 awakenings from 263 participants across 20 published studies --- confirms that dream reports occur in 84.3\% of REM awakenings and 62.6\% of NREM awakenings, with a highly significant stage-dependent association ($\chi^2 = 211.79$, $p = 1.2 \times 10^{-45}$)~[1]. Yet the current computational approach to dream detection treats mentation as a spectral energy problem: state-of-the-art methods extract PSD across six frequency bands and nonlinear statistical moments (catch22), achieving AUC $\approx 0.70$ for REM dream detection~[1]. This ceiling reflects a fundamental limitation: spectral methods measure how much brain activity is present, but not what geometric shape that activity possesses.

We identify a critical dichotomy in neural signal analysis. Statistics and spectral decomposition capture the \emph{energy} of dreaming: power in delta, theta, alpha, sigma, beta, and gamma bands. Topology captures the \emph{shape} of dreaming: the persistent connected components, loops, and voids in the phase-space attractor reconstructed from multi-channel EEG. This distinction is not merely technical. Theoretical frameworks of conscious experience --- including Integrated Information Theory~[2] and Global Workspace Theory~[3] --- predict that dream states involve qualitatively different network integration architectures. Persistent homology provides the mathematical machinery to measure precisely this geometric integration.

Recent TDA work on EEG has demonstrated this approach's power. Baronetzky~[5], located and verified in our most recent reference check, reports 98.16\% sleep-staging accuracy with topological features versus 91.77\% with statistical equivalents. A separate source, Gupta, Beuria \& Behera~[4], reports comparable topological-versus-statistical accuracy gains on meditative-versus-resting-state EEG using persistent-homology and Hodge spectral-entropy features. This source was flagged as unlocatable in an earlier reference check because it had been miscited with an incorrect title, incorrect author initials, and an incorrect volume/page; it has since been re-verified under its correct bibliographic details (Section~5.4) and is treated as corroborating evidence alongside Baronetzky~[5], not as a standalone load-bearing benchmark, since neither source examines dream mentation specifically. Neither study examined dream content discrimination within sleep stages. Simultaneously, the PHINN framework~[6] established topology-conditioned flow matching for rare-event synthesis in financial and epidemiological time series. PHINN-EEG adapts this framework to multi-channel neural signals for the first time.

\paragraph{Contributions.} (i) First topological classification framework for dream content on the DREAM database, targeting AUC $= 0.82$--$0.90$ on the 1{,}462-epoch open-access subset. (ii) Formalisation of Dynamic Betti Curves from multi-channel Takens delay embeddings as discriminative features for dream versus dreamless EEG. (iii) Topology-conditioned rectified flow model for dream-state EEG synthesis, evaluated against spectral-conditioned and unconditional baselines. (iv) A set of candidate Betti transition patterns linking topology to phenomenological dream categories, presented as an exploratory hypothesis space pending empirical validation.

% ======================================================
\section{Related Work}

\subsection{Dream EEG Databases and Classification}

The DREAM database (Wong et al., 2025) is the most comprehensive open-access resource for EEG-based dream research, aggregating 20 independent published studies with 263 participants and 3{,}191 awakenings~[1]. Of these, 2{,}812 awakenings have complete sleep-stage labels (used to establish the data inclusion flow in Section~5.1); 1{,}462 awakenings have publicly available raw EDF files and constitute the open-access subset used for model training. Figure~1 is a schematic illustration only --- no DREAM data were analysed to produce it. The benchmark study establishes AUC $= 0.586$ (NREM) and AUC $= 0.700$ (REM) using normalised PSD across six bands and nonlinear catch22 features.

Moctezuma et al.~[7] extended the DREAM database analysis using Common Spatial Pattern (CSP) features combined with discrete wavelet transform and $k$-nearest-neighbour classification, finding that 8--10 EEG channels suffice for AUROC $> 0.85$. SleepEEGNet~[8] demonstrated 84.3\% overall accuracy and 88.7\% REM-specific recall on the Sleep-EDF-2013 dataset using a single EEG channel (Fpz-Cz), resolving sleep staging but not dream content within stages.

\subsection{Topological Data Analysis in Neuroscience}

Persistent homology has emerged as a powerful tool for neural signal analysis. Giusti et al.~[9] demonstrated that neural population codes exhibit topological structure reflecting stimulus geometry. Lee et al.~[10] identified persistent $\beta_1$ features in fMRI resting-state networks corresponding to known functional connectivity modules. Most directly relevant are Baronetzky~[5] and Gupta et al.~[4] (both verified; see References); neither examined dream mentation. The DREAM database has never been analysed with topological methods. A recent review of TDA methods in EEG signal processing, Ling, Phang \& Liew~[11], surveys this broader literature but likewise does not address dream mentation specifically; this review was initially flagged as unlocatable because it had been miscited under the wrong journal, year, and author name, and has since been re-verified under its correct bibliographic details (Section~5.4).

\subsection{Generative Models for EEG}

A latent-diffusion approach to synthetic sleep EEG generation has been reported by Aristimunha et al.~[12], who use latent diffusion models with a spectral loss term to generate 30-second synthetic sleep-EEG windows. This source was initially flagged as unlocatable because it had been miscited under the wrong venue (IEEE EMBC 2023 rather than its actual venue, the NeurIPS 2023 DGM4H workshop); it has since been re-verified under its correct bibliographic details (Section~5.4). We note it here as related generative work but do not treat its reported results as a performance benchmark for PHINN-EEG, since it targets sleep-stage EEG generation broadly rather than dream-state synthesis specifically. EEG-GAN variants~[22,23] synthesise electrode-resolved signals matching spectral and coherence statistics but suffer from mode collapse and lack topological conditioning. VAEs offer stable training but blur high-frequency phase structure above 20~Hz. WGAN-GP architectures mitigate GAN instability and have been applied to motor imagery and P300 augmentation.

Our topology-conditioned rectified flow offers three advantages: (1) straight optimal-transport paths avoid mode collapse; (2) a topological fidelity metric $L_{\text{topo}}$ (persistence-landscape distance and Euler characteristic matching), computed offline as a validation and conditioning-fidelity target rather than a differentiable training loss (see Section~4.4 for the full clarification of its role); (3) single-step ODE inference achieves $< 500$~ms per epoch versus multiple reverse-diffusion steps. Flow Matching~[13] provides our generative backbone.

% ======================================================
\section{Theoretical Background}

\subsection{Takens Delay Embedding for EEG}

For a single-channel EEG signal $x(t)$ sampled at frequency $f_s$, the Takens delay embedding with dimension $d$ and lag $\tau$ constructs:
\begin{equation}
\Phi_{d,\tau}(x)_t = (x_t, x_{t+\tau}, \dots, x_{t+(d-1)\tau}) \in \mathbb{R}^d
\end{equation}

By Takens' theorem~[14], for generic smooth dynamical systems and appropriate $(d,\tau)$, this embedding diffeomorphically reconstructs the topology of the underlying attractor for a single observable, provided $d \geq 2D+1$, where $D$ is the box-counting (or similarly defined) dimension of the underlying attractor; this is a sufficient, not merely heuristic, condition, and we adopt it explicitly as the basis for dimension selection below.

\paragraph{Dimension selection, corrected for under-embedding risk.} An earlier draft fixed $d=3$ globally, which the reviewer correctly noted is likely insufficient: the intrinsic dimensionality $D$ of EEG dynamics is widely reported to exceed 1 substantially --- commonly $D>4$ even in highly synchronised states --- so the sufficient condition $d \geq 2D+1$ established above can require $d \geq 9$ or higher in the worst case, well above $d=3$. An embedding dimension that is too low relative to the true attractor dimension risks projection-induced self-intersections of the reconstructed trajectory, which can manufacture spurious loops and voids that do not exist in the true attractor and would confound the persistent-homology features this pipeline relies on. We therefore revise the primary pipeline as follows. Dimension $d$ and lag $\tau$ are still determined once, prior to the pre-registered pipeline, via a pilot false nearest-neighbour (FNN) analysis and autocorrelation zero-crossing analysis on a representative sample of channels and epochs (not re-estimated per epoch); $\tau$ is set to the pilot-estimated first zero-crossing of the autocorrelation function (reported per-dataset in the released protocol appendix). For $d$, rather than fixing a single global value a priori, we pre-register a candidate set $d \in \{5,7,10,15\}$, spanning the practically-estimated FNN range up through values consistent with the higher end of plausible EEG attractor dimensionality. All primary classification results (Model~1) are computed and reported for each candidate $d$ in this set as a pre-specified sensitivity analysis; the single default value used for the primary headline results is $d=7$, selected via nested cross-validation on the training partition only (never using test-set performance), locked prior to any test-set evaluation. If classification performance degrades sharply at higher $d$ within the sweep, this would be reported as evidence that lower-$d$ results benefit from under-embedding artefacts rather than genuine topological signal, per the reviewer's concern. No per-epoch or per-channel dynamic re-estimation of $d$ is performed within any single candidate configuration in the primary pipeline.

\paragraph{Multivariate extension (note).} For multi-channel EEG with $K$ channels, we construct a joint embedding by concatenating per-channel delay vectors at each time step $s$ into a point in $\mathbb{R}^{K \times d}$. This construction does not automatically inherit the diffeomorphic reconstruction guarantees of the univariate theorem --- multivariate extensions require that the combined observation functions span the attractor, which is not guaranteed by concatenation alone~[20,21]. We therefore treat the joint embedding as a practical heuristic that captures cross-channel phase relationships, with theoretical justification drawn from the single-channel case for each channel individually. We make no claim of diffeomorphic reconstruction for the concatenated space.

Because this concatenation does not guarantee a diffeomorphic embedding, the geometric features extracted from the joint embedding should be interpreted as descriptors of cross-channel phase-relationship geometry rather than as a direct reconstruction of a true low-dimensional neural attractor; the Geometric Dream Hypothesis (Section~3.3) is correspondingly framed as a hypothesis about this joint-embedding geometry, not about the ground-truth neural state-space topology, and this caveat applies wherever attractor-level language is used elsewhere in this paper. The heuristic draws on the general property that joint delay embeddings can capture cross-channel phase geometry invisible to PSD and coherence measures, as reported by Baronetzky~[5] and Gupta et al.~[4] (both verified; Section~5.4); per the reviewer's concern (Section~5.4), we do not claim these sources specifically validate our concatenated joint-embedding construction. To be precise about precedent: neither~[4] nor~[5] applies persistent homology to a concatenated multi-channel delay embedding of the specific form used here ($\mathbb{R}^{K \times d}$ formed by stacking per-channel delay vectors); [4] and [5] instead apply topological methods within a single channel or to functional-connectivity graphs derived from cross-correlation matrices, which is the more standard TDA-EEG construction. We therefore explicitly acknowledge that the concatenated joint embedding used in this paper is a novel heuristic without direct empirical precedent in the cited literature, rather than an established technique demonstrated by prior work; Sections~3.1 and~6.1 already caveat its interpretation accordingly, and Section~4.6(v)'s channel-perturbation control is our primary empirical check on whether this heuristic is capturing genuine multi-channel structure.

We considered two established alternatives to simple concatenation and did not adopt them for this pipeline, for the following reasons, which we flag as a methodological limitation rather than a settled justification: (a) Deyle \& Sugihara-style multivariate/multi-view state-space reconstruction~[20], which selects or weights observation channels according to their individual reconstruction quality (e.g., via simplex-projection cross-validation) rather than concatenating all channels uniformly, would better respect channels of differing signal quality but requires a per-channel quality-selection step we have not yet specified or validated for EEG; (b) a PCA- or dynamical-component-analysis-based joint embedding, which projects the multi-channel signal onto a lower-dimensional set of components before delay-embedding, would reduce the ambient dimension (mitigating the curse-of-dimensionality concerns raised in Section~6.3) but introduces its own diffeomorphism-preservation questions and an additional hyperparameter (component count) that we have not characterised. Uniform concatenation was chosen over both for implementation simplicity and because it makes no channel-selection or dimensionality-reduction assumptions beyond what is already flagged above.

Rather than deferring this comparison entirely to future work, we pre-register a secondary sensitivity analysis: on the harmonised-reference datasets, we additionally construct a PCA-projected joint embedding (retaining components explaining 95\% of variance, fit per-fold on the training partition only) as an alternative to uniform concatenation, recompute the Dynamic Betti Curves and downstream classification AUC (Model~1) from this alternative embedding, and report the resulting AUC alongside the primary concatenation-based result. Substantial degradation in discriminative power under concatenation relative to the PCA-projected embedding would indicate that the uniform-concatenation heuristic is introducing spurious geometric artefacts rather than capturing genuine multivariate structure; comparable performance across both embeddings is treated as descriptive corroboration, not proof, that the simpler heuristic is not the source of the reported effect. Adopting the Deyle \& Sugihara-style channel-weighting alternative remains identified as a priority direction for future empirical follow-up work, since it requires a per-channel quality-selection step we have not yet specified or validated for EEG. The point cloud $PC_t$ contains $W$ points in $\mathbb{R}^{K \times d}$, capturing inter-channel phase relationships through their joint geometry. Critically, $\beta_2$ is computed from this delay-embedded cloud, not from the $K$ physical electrode positions themselves.

We further note that the joint embedding is subject to volume conduction: because scalp electrodes simultaneously record instantaneous mixtures of the same underlying cortical sources, zero-lag spatial correlations between channels are present in the raw signal itself, independent of any true dynamical phase relationship. Consequently, topological features extracted from the concatenated point cloud may partly reflect this spatial mixing rather than genuine cross-channel dynamical coupling. A spherical-spline surface Laplacian can in principle target this zero-lag mixing directly, but it requires a dense, spatially distributed unipolar montage (typically $\geq 32$ channels) to be well-posed; because the open-access DREAM subsets used here are sparse (approximately 6--18 channels), we do not apply it to the primary pipeline (Section~4.1 gives the full montage-density criterion and rationale). Volume conduction therefore remains an uncorrected structural limitation of the multivariate construction for all open-access subsets --- unipolar and bipolar alike --- and its magnitude is assessed only indirectly, via the surrogate-data (Section~4.6(iv)) and channel-perturbation (Section~4.6(v)) controls.

We revise a previous version of this note: an earlier draft used univariate (per-channel) IAAFT surrogates here, which preserve each channel's own linear autocorrelation and amplitude distribution but destroy cross-channel linear correlation entirely, and therefore could not dismantle zero-lag spatial mixing (volume conduction) while leaving it otherwise intact --- since volume conduction is itself a zero-lag linear cross-channel effect, univariate surrogates lacked the confound they were meant to test for. We now use MIAAFT surrogates (Section~4.6(iv)), which jointly preserve the cross-channel cross-spectral density matrix, and so do retain volume-conduction-driven zero-lag mixing while destroying genuine nonlinear (within- and cross-channel) structure; a real-vs-surrogate discrimination under MIAAFT therefore constitutes a meaningful, though still indirect, test of whether extracted topological features reflect genuine nonlinear coupling beyond what linear cross-channel mixing (including volume conduction) can produce. Because the spherical-spline surface Laplacian cannot be validly applied to the sparse montages used here (Section~4.1), it is not available as an additional check for either unipolar or bipolar subsets in the primary pipeline. The surrogate control (Section~4.6(iv)) and the channel-perturbation control (Section~4.6(v)) therefore remain the only available checks across all open-access subsets, and the residual volume-conduction confound is recorded as an open methodological limitation for the full dataset, not only for a subset of it (Section~6.5).

\subsection{Persistent Homology and Betti Curves}

Given point cloud $PC_t$, we build the Vietoris--Rips filtration:
\begin{equation}
VR(PC, \varepsilon) = \{ \sigma \subseteq PC : \mathrm{diam}(\sigma) \leq \varepsilon \}
\end{equation}

Persistent homology tracks the birth and death of topological features across the filtration, yielding persistence diagrams $PD_0, PD_1, PD_2$ recording (birth, death) pairs for connected components ($H_0$), loops ($H_1$), and voids ($H_2$). The stability theorem~[24,25] yields:
\begin{equation}
d_B(PD_k(P), PD_k(Q)) \leq 2\, d_H(P,Q)
\end{equation}

where the factor of 2 arises directly from the triangle inequality used in constructing interleaving maps between the Vietoris--Rips filtrations of $P$ and $Q$ themselves: if $P$ and $Q$ are $\varepsilon$-close in Hausdorff distance, a simplex of diameter $\delta$ in $P$ maps to a set of diameter at most $\delta + 2\varepsilon$ in $Q$, and this $2\varepsilon$ slack propagates to the bottleneck-distance bound above~[24,25]. (This is distinct from, and should not be conflated with, the separate multiplicative interleaving between Vietoris--Rips and \v{C}ech complexes constructed on the same point cloud, which relates two different complex types rather than the same complex type across two different point clouds.) This bound provides noise robustness sufficient for EEG, where electrode impedance and muscle artefact introduce perturbations of order $\delta \ll \varepsilon^*/2$.

\paragraph{Multi-scale Betti features, with a bounded filtration range (correcting a prior internal inconsistency).} Rather than evaluating Betti numbers at a single fixed scale $\varepsilon^*$, we extract features across a range of filtration scales --- but this range is explicitly bounded, not the theoretically complete filtration out to a fully connected complex, since the latter is what makes the 50th-percentile-threshold complex intractable (below). The 10th-percentile threshold itself is not asserted on theoretical grounds alone: it is selected via a pre-registered percentile sweep (5th, 10th, 15th, 20th percentiles) evaluated only on each LODO fold's training partition, choosing the percentile that maximises training-fold classification AUC before the held-out fold is touched, so the final scale is data-driven while remaining leakage-free; the 10th percentile is reported as the value expected, based on pilot exploration, to be selected by this procedure, with the full sweep results reported descriptively alongside the primary result to show how sensitive the Betti curves are to this choice. Concretely: Dynamic Betti Curves $\beta_k(t)$ are computed as the Betti number at the single filtration value $\varepsilon^*(t)$, set at the 10th percentile of pairwise distances within each individual sliding sub-window's point cloud (the fixed-size, 500-sample sub-window centred at time $t$ --- not the full 30-second epoch; sub-window-specific). Separately, the persistence landscapes and persistent entropy described below (feature category (iii)) are computed from a birth-death diagram whose filtration is capped at a maximum scale $\varepsilon_{\max}(t)$ fixed at the 25th percentile of the same sub-window's pairwise distances --- bounded well below the 50th-percentile regime, not extended to the fully connected complex. This $\varepsilon_{\max}$ cap means some longer-lived topological features that would only die beyond the 25th-percentile scale are right-censored (recorded as persisting to $\varepsilon_{\max}$ rather than to their true death scale); we treat this as an explicit, pre-registered approximation to the persistence diagram --- a bounded-scale diagram, not the unbounded ``full'' diagram --- and note this distinction to avoid the earlier draft's inconsistency between claiming a computationally intractable full diagram while also asserting the 50th-percentile threshold alone was already intractable. Setting the primary Betti-curve threshold at the 10th percentile, and the landscape/entropy cap at the 25th percentile, means the complex explored remains well short of the dense, tens-of-millions-of-simplices regime that a full 50th-percentile-or-beyond filtration would require for $n=500$ points --- the regime we explicitly avoid as computationally intractable for $H_2$ computation (Section~6.3). A 90th-percentile threshold, used in an even earlier draft, produced nearly-complete complexes and is likewise avoided. Both $\varepsilon^*(t)$ and $\varepsilon_{\max}(t)$ are computed from each sub-window's own point cloud, not from the full epoch and not globally, so that the bounded filtration range tracks local point-cloud density rather than a fixed global scale.

We additionally extract persistence landscapes $\lambda_{kn}(t)$ and persistent entropy $H_{\text{pers}}$ as multi-scale summaries of the bounded-range persistence diagram described above (filtration capped at $\varepsilon_{\max}(t)$, the sub-window's own 25th-percentile pairwise distance), ensuring discriminative signal is captured across a range of scales rather than at the single $\varepsilon^*(t)$ value alone, while remaining within the tractable filtration range established above.

\subsection{The Geometric Dream Hypothesis}

We formalise the central hypothesis. Define the Euler characteristic trajectory:
\begin{equation}
\chi(t) = \beta_0(t) - \beta_1(t) + \beta_2(t)
\end{equation}

\paragraph{Pre-registered hypotheses.}

\textbf{H1 (Classification):} PHINN-EEG topological features achieve AUC $= 0.82$--$0.90$ on REM dream detection. The primary confirmatory test computes AUC independently within each of the harmonised-reference held-out datasets (Section~6.2) for both Model~1 (topological features) and Model~2 (PSD/catch22 features; Section~4.3) --- the two being evaluated on the identical per-dataset LODO folds --- macro-averages the per-dataset AUCs for each model (unweighted mean across datasets), and compares them via a paired one-sample $t$-test on the per-dataset AUC differences (Model~1 minus Model~2, two-sided, $p<0.05$), since paired evaluation on matched folds is required to attribute any difference to the feature set rather than to fold-specific difficulty. We use two-sided tests throughout (here and in H2, H3) rather than one-sided, since the topological feature set is novel and its efficacy on this specific task is unproven --- it is not a theoretical impossibility for Model~1 to underperform Model~2, so a one-sided test would fail to control the Type~I error rate against an unexpected reversal in either direction. The externally published Wong catch22 benchmark (AUC $= 0.700$), whose reported performance was established under a different evaluation protocol and fold structure than the one used here, is reported alongside as descriptive context rather than as the basis for a formal statistical test, since comparing this study's fold-level variance against a point estimate derived from a different scheme would confound genuine model superiority with fold-selection difficulty. Given the small number of datasets (4 unipolar, plus the bipolar subsets handled separately per Section~6.2), a non-parametric sign test lacks the power to serve as a co-equal decision criterion: for the 4-dataset unipolar group, the minimum attainable two-sided sign-test $p$-value is 0.125 (0.0625 one-sided), which cannot cross the $p<0.05$ threshold regardless of the true effect size. We therefore report the exact two-sided sign-test $p$-value as a descriptive, directional robustness indicator rather than as a mandatory co-equal test, and treat H1 as supported on the basis of the paired $t$-test against Model~2, with sign-test agreement (where the dataset count permits significance) and clearance of the descriptive 0.700 benchmark reported as corroborating evidence. Because a parametric $t$-test on 4--6 macro-averaged data points cannot reliably assess its own normality assumption, we additionally report, as a further robustness check not conditioning the primary decision, a generalised linear mixed-effects model (GLMM) fit on epoch-level binary predictions (correct/incorrect at the operating threshold) with model identity (Model~1 vs Model~2) and sleep stage as fixed effects and participant nested within dataset as a random intercept, testing the Model~1 vs Model~2 fixed-effect coefficient; this approach uses the full epoch-level sample size rather than reducing the comparison to a handful of macro-averaged points, and its agreement or disagreement with the paired $t$-test is reported descriptively alongside the primary result. We deliberately do not pool raw predictions across datasets before computing AUC, since datasets differ in their base rate of dream recall and pooled-prediction AUC is vulnerable to Simpson's-paradox-style artefacts (a model can achieve a high pooled AUC by exploiting between-dataset base-rate differences alone, without genuine within-dataset discriminative power); macro-averaging per-dataset AUCs avoids this. Per-dataset AUCs are additionally reported descriptively (Section~5.3) to characterise inter-laboratory heterogeneity.

\textbf{H2 (NREM):} Topological features achieve AUC $= 0.72$--$0.78$ on NREM dream detection (vs AUC $= 0.586$ baseline), tested via the same macro-averaging approach as H1.

\textbf{H3 (Ablation):} PCA-reduced Model~1 (topology, XGBoost, reduced to Model~2's native $\approx$220-dimensional budget via per-fold training-only PCA; Section~5.3) significantly outperforms native Model~2 (raw EEG, XGBoost; Section~4.3) on macro-averaged REM AUC, using the same per-dataset-then-average procedure and the same paired one-sample $t$-test (on the per-dataset AUC differences between PCA-reduced Model~1 and Model~2, two-sided, $p<0.05$) plus sign-test robustness check specified for H1 above, since the two are evaluated on the same per-dataset folds and so yield paired, not independent, AUC estimates; the Model~1 versus Model~2b comparison is retained and reported descriptively for continuity with the original protocol.

\textbf{H4 (Generation):} Topology-conditioned flow achieves SCR $>30\%$ (vs 3--5\% random baseline) and lower FED than the unconditional, spectral-conditioned, EEG-GAN, VAE, and diffusion baselines, with statistical significance assessed via the bootstrap procedure specified in Section~5.3.

The primary family-wise error rate for H1--H3 is controlled via Bonferroni correction across all hypothesis-metric combinations tested ($m=15$: 3 hypotheses $\times$ 5 metrics; Section~5.3), not per metric alone.

\paragraph{Hypothesis (Dream Topology).} Dream states exhibit stable, non-trivial topological structure --- characterised by sustained $\beta_1 > 0$ (persistent loops, here interpreted as a geometric correlate of recurrent neural activity, though we explicitly note this mapping is a theoretical conjecture requiring empirical validation and does not imply direct correspondence with macroscopic brain network loops) and moderate $\beta_0$ fragmentation --- while dreamless states collapse toward trivial topology characterised by high $\beta_0$ and $\beta_1 \approx 0$. Figure~1 provides a schematic illustration of this predicted divergence. This directional prior is not treated as a safe default: dreamless NREM epochs are often characterised by highly synchronised, high-amplitude slow oscillations and sleep spindles, and such strongly periodic or quasi-periodic dynamics can themselves form robust limit cycles in a delay-embedded phase space, yielding sustained $\beta_1 > 0$ independent of dream content. Equating phenomenological ``emptiness'' with an absence of topological loops would conflate reportable dream content with the underlying structure of the dynamics, which is not necessarily trivial during unconscious sleep stages. We therefore treat the trivial-topology prediction for dreamless states as an empirically testable component of H1/H2 (Section~3.3), evaluated within matched sleep stages (dream vs. dreamless awakenings drawn from the same stage) rather than assumed, and a result in which dreamless epochs instead show sustained $\beta_1 > 0$ driven by slow-oscillation or spindle activity is treated as informative about the mechanism rather than as invalidating the classification framework, since classification (H1/H2) depends only on topological features discriminating dream from dreamless epochs, not on the trivial-topology direction of that discrimination holding as predicted.

\paragraph{Phenomenological caveat.} The mapping from topological invariants in delay-embedded phase space to phenomenological dream content (e.g., equating $\beta_1$ loops with ``nightmares'' or ``recurrent neural activity'') is speculative. The Takens embedding of multi-channel EEG produces an abstract high-dimensional phase space; geometric features therein are not straightforwardly analogous to macroscopic brain network loops. All candidate pattern labels are theoretically motivated priors to be tested empirically --- they are not established correspondences and are not part of any finished classification scheme.

\begin{figure}[htbp]
\centering
\includegraphics[width=0.95\textwidth]{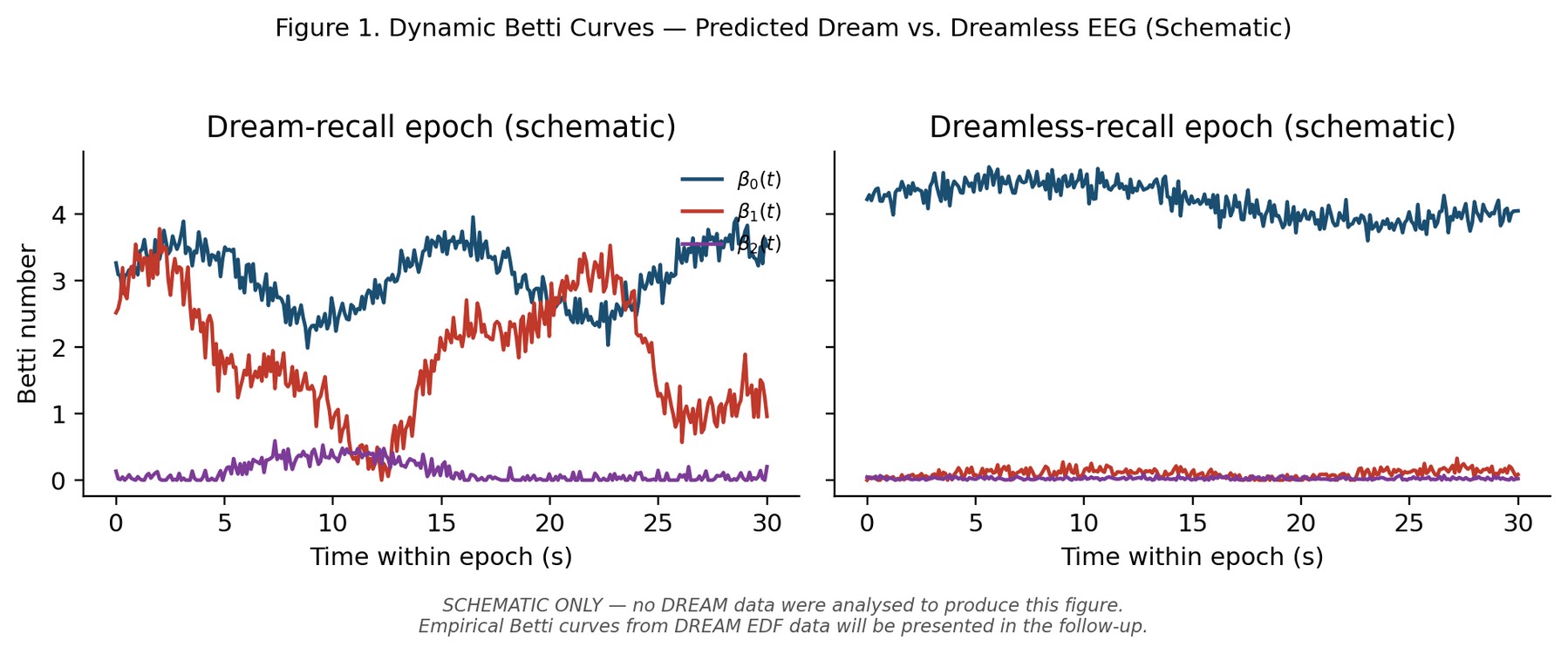}
\caption{Dynamic Betti Curves --- Predicted Dream vs. Dreamless EEG (Schematic). SCHEMATIC ONLY --- no DREAM data were analysed to produce this figure. Empirical Betti curves from DREAM EDF data will be presented in the follow-up.}
\end{figure}

% ======================================================
\section{Methodology: The PHINN-EEG Pipeline}

\subsection{Data Architecture and Preprocessing}

We use the open-access subsets of the DREAM database for which raw .edf signal files are publicly available: Lacaux (N1 imagery, 2021), Zhang and Wamsley (2019), Scarpelli~1 and~2 (2019--2020), Siclari (2017), and Sikka (2018) --- constituting 1{,}462 labelled awakening epochs from 201 of the 263 unique participants in the full DREAM registry (the open-access subsets do not cover all 263 participants; the exact per-subset participant count will be confirmed and reported in the CONSORT-style flow diagram, Section~5.1). For sleep-staging validation (without dream labels), we use the Sleep-EDF Expanded database (PhysioNet): 197 whole-night recordings, expert-scored~[15].

Dream report labels are binarised: 1 = vivid or structured narrative recall; 0 = no experience or vague imagery, following Wong et al.'s primary analysis. EDF files are loaded via MNE-Python~[16]. Where unipolar/raw-referenced recordings are available, all channels are re-referenced to Cz (or linked mastoids where Cz is absent across that subset) prior to any further processing. Where a dataset provides only pre-computed bipolar derivations (e.g., Fpz-Cz, C4-M1), re-referencing to Cz is not attempted; the pipeline instead operates directly on the bipolar signal as a differential measure for that dataset (full protocol and rationale in Section~6.2).

We then apply a zero-phase (forward-backward, \texttt{filtfilt}) 4th-order Butterworth bandpass filter (0.5--45~Hz), a 50~Hz notch filter, and robust $z$-score normalisation per channel per recording. Zero-phase filtering is used specifically because Takens delay embedding reconstructs attractor geometry from the precise phase relationships between time-lagged samples; a causal (forward-only) filter would introduce phase distortion that could alter the resulting topological features. Bad channels are identified via RANSAC interpolation. Epochs are defined as 30-second windows immediately preceding each awakening.

\paragraph{Spatial control for volume conduction, gated by montage density.} The spherical-spline surface Laplacian (current source density) is a well-validated volume-conduction control, but it requires a sufficiently dense, spatially distributed unipolar montage (typically $\geq 32$ channels) to estimate the spatial second derivative accurately; applying it to sparse montages produces ill-posed spline fits and can introduce severe artefacts rather than removing them. The open-access DREAM subsets are, in practice, sparse: unipolar-referenced subsets in our inclusion set range from approximately 6 to 18 channels, well below this threshold. We therefore do NOT apply the full spherical-spline surface Laplacian to any of the primary open-access subsets, correcting the invalid blanket application in an earlier draft. Instead, preprocessing for unipolar subsets proceeds as in Section~3.1/4.1 without a spatial Laplacian transform, and the residual volume-conduction confound for these subsets is addressed only via the MIAAFT surrogate control (Section~4.6(iv)) and channel-perturbation control (Section~4.6(v)), consistent with the discussion in Section~3.1. Should a future open-access DREAM release, or the Sleep-EDF validation set, provide a sufficiently dense montage ($\geq 32$ channels) for a given subset, the spherical-spline surface Laplacian will be applied to that subset specifically as a pre-specified, montage-gated secondary sensitivity analysis; it is not part of the primary pipeline for the current, sparse-montage open-access subsets.

\paragraph{Subject-level data partitioning.} All cross-validation splits are grouped at the participant level --- no subject's epochs appear in both training and test partitions. This is enforced via subject-stratified GroupKFold in both the 5-fold tuning loop and leave-one-dataset-out evaluation.

\paragraph{Class imbalance handling (pre-specified primary method).} For Stage~1 pre-registration, class-weighted XGBoost loss is designated as the sole pre-specified primary analysis method for testing H1 and H2. SMOTE oversampling constitutes a pre-specified secondary ablation only, applied to a PCA-reduced projection of the topological feature vector (retaining 95\% variance) where linear distances better approximate local geometry. We explicitly acknowledge that PCA projection does not fully resolve non-Euclidean manifold constraints --- synthetic SMOTE samples may still violate topological validity --- which is the principal reason class-weighted loss is preferred as the primary method. The SMOTE-in-PCA-space ablation's results will be reported descriptively in Stage~2 but will not influence the primary hypothesis tests, eliminating researcher degrees of freedom. This rule is locked prior to any test-set observation.

\subsection{Topological Feature Extraction Engine}

Our pipeline implements four stages.

\textbf{(i) Delay Embedding:} For each EEG channel $k$ and epoch, we apply the pilot-estimated $\tau$ established in Section~3.1 together with the default embedding dimension $d=7$ (FNN and autocorrelation analysis is performed once, at protocol-design time, not per-epoch); the full pre-registered dimension sensitivity sweep additionally re-runs this stage at $d \in \{5,10,15\}$ (Section~3.1). Construct per-channel delay vectors and assemble joint point cloud $PC_t$ using a fixed sliding sub-window: window $W = 500$ samples (5~s at 100~Hz), stride $= 100$ samples (1~s), overlap $= 80\%$. These parameters are fixed for all epochs and datasets and are pre-registered as the primary configuration; no dynamic or signal-dependent resizing of the window is performed. We flag as a limitation that a 5-second window is short relative to the period of delta-band activity (0.5--4~Hz), completing as few as 2.5 cycles at the lower bound of that range; reconstructing a phase-space attractor at the pre-registered embedding dimensions ($d=7$--15) from so few cycles of a low-frequency component risks a sparse, unstable manifold whose Betti numbers partly reflect windowing artefacts rather than settled underlying geometry. We do not resolve this within the primary pre-registered configuration, but the window-overlap sensitivity analysis above (Section~6.5) and a planned frequency-band-stratified windowing scheme --- using longer sub-windows for slow-wave-dominant epochs and the current window for faster activity --- are identified as a priority follow-up to disambiguate genuine low-frequency topology from window-length artefacts.

\paragraph{Pre-specified sensitivity analyses (window overlap).} Model~1 classification performance (REM AUC) is additionally evaluated at three alternative stride settings --- stride $=175$ samples (65\% overlap), stride $=250$ samples (50\% overlap), and stride $=500$ samples (0\% overlap, i.e., strictly non-overlapping sub-windows) --- with all other pipeline parameters held fixed, to characterise how much of the classification signal depends on the 80\%-overlap-induced autocorrelation described in Section~6.5. The non-overlapping (0\%) setting is included specifically as a baseline that is structurally immune to the overlap-induced autocorrelation concern, since consecutive sub-windows share no samples; a large drop in AUC at lower overlap, and in particular at the non-overlapping baseline, would indicate the primary 80\% setting is exploiting overlap-induced autocorrelation rather than genuine dream-related topology. Results across all four overlap settings are reported together.

\paragraph{Pre-specified sensitivity analysis (adaptive vs. fixed filtration scale).} Because the adaptive, per-sub-window 10th-percentile filtration scale $\varepsilon^*(t)$ (Section~3.2) normalises for local point-cloud density, it could in principle produce similar Betti curves for a high-amplitude structured signal and a low-amplitude noise process whose relative internal geometry happens to be similar, since both are rescaled to fill the same normalised space. To test whether this adaptive normalisation is itself manufacturing classification signal, we additionally run the full topological feature extraction pipeline using a fixed filtration scale calibrated separately within each LODO fold (set once per fold, from the pooled pairwise-distance distribution across a representative pilot sample of sub-windows drawn only from that fold's training datasets, excluding the held-out dataset, and then held constant for every sub-window and every epoch within that fold) as a parallel, pre-specified comparison. This per-fold recalibration is necessary precisely because a single scale pooled across all subsets --- as in an earlier draft of this control --- would include the held-out dataset for every fold, leaking the test set's distance distribution into the feature-extraction pipeline ahead of evaluation and invalidating the sensitivity analysis; calibrating exclusively from each fold's training partition avoids this leakage. Model~1 classification performance is reported for both the adaptive-scale (primary) and fixed-scale (parallel) pipelines; if the fixed-scale pipeline achieves comparable REM AUC to the adaptive-scale pipeline, this would support the interpretation that genuine topological structure --- not adaptive-scale normalisation artefacts --- drives classification performance, and a large discrepancy in the opposite direction would be reported as evidence that the adaptive scale is inflating apparent performance.

\textbf{(ii) VR Filtration:} We use Ripser~[17] for fast Vietoris--Rips computation, run with its maximum filtration radius explicitly capped at $\varepsilon_{\max}(t)$ (Section~3.2) rather than to completion, keeping every filtration within the tractable regime established there. Filtration is performed independently on each sliding sub-window's point cloud (stage (i) above), not on the full 30-second epoch as a single object. For a dataset with $K$ channels ($K$ ranges from approximately 6 to 18 across the primary open-access subsets; Section~4.1) and the default embedding dimension $d=7$ (Section~3.1; see below for the dimension sensitivity sweep), each 500-sample sub-window yields a joint point cloud of up to $\approx 500$ points in $\mathbb{R}^{K \cdot d}$; this ambient dimension scales both with $K$ (fixed per dataset) and with the candidate embedding dimension used across the sensitivity sweep ($d = 5,7,10,$ or $15$). For illustration, an 8-channel subset (matching the channel count used in Moctezuma et al.~[7]) at $d=7$ gives an ambient dimension of $\mathbb{R}^{56}$, and at $d=5,10,$ or $15$ gives $\mathbb{R}^{40}$, $\mathbb{R}^{80}$, or $\mathbb{R}^{120}$ respectively; the same formula applies at each dataset's own $K$. Concatenating delay vectors across the full 30-second epoch (3{,}000 samples at 100~Hz) before windowing would instead yield a substantially larger $\approx 3{,}000$-point cloud in the same ambient dimension --- we never construct or filter that larger object. To bound sub-window point-cloud size (which can exceed 500 points near sub-window edges), we apply landmark-based sub-sampling (farthest-point sampling, maxmin landmark selection) within each sub-window to cap it at $n=500$ points; for a native 500-sample sub-window this is close to a no-op and serves primarily as a safety bound rather than an active reduction. Because Vietoris--Rips complexity for a fixed $n$ is governed by the number of points and the filtration cap rather than by the ambient dimension directly, increasing $d$ within the sensitivity sweep does not itself change the $O(n^2)/O(n^3)/O(n^4)$ simplex-count scaling below, though it does increase the per-pairwise-distance computation cost linearly in $d$. The characteristic scale $\varepsilon^*(t)$ is set at the 10th percentile of pairwise distances within each individual sub-window's point cloud (see Section~3.2). The scale is computed per sub-window, from that sub-window's own point cloud --- not from the full 30-second epoch and not globally. Computing $\beta_0$ alone requires only the 1-skeleton (vertices and edges), since connected components are determined entirely by the graph structure without reference to any higher simplices. $\beta_1$, by contrast, cannot be recovered from the 1-skeleton alone: identifying which 1-cycles are non-trivial loops (rather than boundaries of filled-in 2-dimensional faces) requires knowing the 2-skeleton, i.e., which triangles (2-simplices) are present in the complex, so we extend computation to the 2-skeleton when computing $\beta_1$. We further extend to the 3-skeleton (tetrahedra, i.e., 3-simplices) when computing $\beta_2$, since $H_2$ similarly cannot be recovered from boundary maps defined only up to the 2-skeleton --- a non-trivial $\beta_2$ requires the absence of the 3-simplices that would otherwise bound (and thereby trivialise) a given 2-dimensional void; the 3-skeleton is computed precisely so that voids filled by 3-simplices can be identified and excluded, isolating the 2-cycles that remain genuinely non-trivial. This yields complexity of $O(n^2)$ edges for $\beta_0$; an additional $O(n^3)$ 2-simplices when $\beta_1$ is computed; and a further $O(n^4)$ 3-simplices when $\beta_2$ is computed, per sub-window; a full 30-second epoch comprises approximately 26 overlapping sub-windows given the stride $=100$-sample grid, so total epoch-level cost scales linearly with this per-window cost (see Section~6.3 for the resulting timing estimate). Homology is computed to dimension 2 ($H_0, H_1, H_2$).

\textbf{(iii) Feature vector:} Per epoch we extract: Dynamic Betti Curves $\beta_0, \beta_1, \beta_2$ across the sliding sub-window grid ($W=500$ samples, stride $=100$ samples); Persistence Landscapes $\lambda_{kn}(t)$ for $k=1,2$, $n=1\ldots5$; Persistent Entropy $H_{\text{pers}}$; Euler Characteristic $\chi(t)$; and the Certified Persistence Ratio (CPR: the fraction of topological features whose persistence exceeds a bootstrap noise threshold, quantifying robustness to point cloud perturbation). Total feature dimensionality: $\approx 340$ real-valued features per epoch.

\subsection{Classification Architecture}

We train classifiers on the topological feature vectors, designed to enable a valid ablation of topological contribution while holding architecture (and, where applicable, feature-set dimensionality) constant.

\paragraph{Model 1 --- Topology Ensemble (XGBoost):} XGBoost~[18] on the full topological feature vector ($\approx 340$ features). Class imbalance handled via primary method (class-weighted loss, Section~4.1).

Hyperparameters are tuned via 5-fold GroupKFold within the training set only, using Bayesian optimisation (Optuna, Tree-structured Parzen Estimator, 100 trials) over the following pre-specified search space: \texttt{max\_depth} $\in [3,10]$; \texttt{learning\_rate} $\in [0.01, 0.3]$ (log-uniform); \texttt{n\_estimators} $\in [100, 1000]$; \texttt{subsample} $\in [0.5, 1.0]$; \texttt{colsample\_bytree} $\in [0.5, 1.0]$; \texttt{min\_child\_weight} $\in [1, 10]$. The selection metric is mean validation AUC across the 5 GroupKFold folds; the configuration maximising this metric is retained and refit on the full training partition before evaluation on the held-out LODO dataset. This search space and procedure are identical across Model~1, Model~2, and Model~2b (below) to avoid confounding tuning effort with feature-set identity.

\paragraph{Model 2 --- Raw EEG Ensemble (XGBoost, ablation):} XGBoost on raw EEG statistical features (PSD, catch22, band ratios --- $\approx 220$ features) with no topological input. Same architecture as Model~1, different feature set only.

Because Model~1 ($\approx 340$ features) and Model~2 ($\approx 220$ features) differ in feature-set size as well as feature type, comparing them alone confounds the discriminative value of topological features with sheer feature-space capacity. We therefore additionally report Model~2b --- a dimensionality-matched raw-EEG ablation in which the Model~2 feature set is expanded with additional multi-taper spectral bins and higher-order catch22 / statistical-moment features to match the $\approx 340$-feature budget of Model~1, using the same XGBoost architecture and hyperparameter search space. We recognise that padding Model~2 with additional spectral bins risks adding uninformative, noisy dimensions rather than genuinely matched information content, which could artificially depress Model~2b's performance and inflate Model~1's apparent advantage; dimensionality matching is not the same as information matching. As a complementary check that does not rely on feature padding, we additionally apply PCA to the $\approx 340$-dimensional topological feature vector, retaining the top $\approx 220$ components (matching Model~2's native dimensionality), and compare this PCA-reduced Model~1 variant against Model~2 directly. H3 is tested primarily via this PCA-reduced Model~1 versus native Model~2 comparison, since it is the only comparison that does not depend on artificially padding either model's feature set with potentially uninformative dimensions and therefore cannot be biased by feature-padding noise in either direction. The Model~1 versus Model~2b comparison (dimensionality-matched by padding Model~2 instead) is retained and reported alongside as a secondary, descriptive robustness check for continuity with the original protocol, together with the native-dimension Model~1 versus Model~2 result.

\paragraph{Model 3 --- Temporal Fusion (CNN-BiLSTM):} A 1D-CNN followed by a Bidirectional LSTM taking as input both the raw EEG epoch and the topological feature vector via cross-attention:
\begin{align}
h_{\text{topo}} &= \text{FeatureEncoder}(\beta^*)\\
h_{\text{eeg}} &= \text{CNN-BiLSTM}(x_{\text{raw}})\\
\hat{y} &= \text{Softmax}\big(\text{MLP}([h_{\text{eeg}} ; \text{CrossAttn}(h_{\text{eeg}}, h_{\text{topo}})])\big)
\end{align}

\paragraph{Architecture and training details (pre-registered):} CNN branch --- 4 one-dimensional convolutional layers with channel widths $32 \to 64 \to 128 \to 256$, kernel size 5, stride 1, ReLU activations, batch normalisation after each layer, and max-pooling (factor 2) after every second convolutional layer. BiLSTM --- 2 stacked layers, 128 hidden units per direction (256-dimensional combined hidden state). FeatureEncoder for $h_{\text{topo}}$ --- since $\beta^*$ is a single $\approx 340$-dimensional aggregate vector rather than a sequence (Section~4.6), applying multi-head self-attention to it would attend only to itself (sequence length 1) and add no representational power beyond a feed-forward network; $h_{\text{topo}}$ is therefore computed with a 2-layer MLP (hidden dimension 128, GELU activation, LayerNorm after each layer) applied to $\beta^*$, without any attention mechanism. Optimiser: AdamW, learning rate $1 \times 10^{-4}$, weight decay $1 \times 10^{-5}$, cosine learning-rate schedule. Batch size: 32. Dropout: 0.3 applied after each CNN block, each LSTM layer, and each Transformer encoder layer. Stopping criterion: early stopping on validation AUC with patience $=15$ epochs, maximum 200 epochs, best-checkpoint restoration.

Sleep stage is included as a categorical covariate in all models. The PHINN-No-Topo baseline used in prior drafts (CNN-BiLSTM on raw EEG only) is superseded by the cleaner ablation in Model~2/2b; the CNN-BiLSTM without topological input would confound architecture with feature set.

\paragraph{Validation scheme mapping (Table~1).} To resolve the apparent tension between 5-fold CV and LODO CV: the 5-fold GroupKFold loop is used exclusively for hyperparameter tuning within the training partition; LODO across the 6 open-access EDF subsets provides the final generalisation estimate reported for all primary hypotheses.

\begin{table}[htbp]
\centering
\small
\begin{tabular}{p{2.6cm}p{7.4cm}p{3.2cm}}
\toprule
\textbf{Scheme} & \textbf{Purpose} & \textbf{Metric} \\
\midrule
5-fold GroupKFold & Hyperparameter tuning within training set only & Dev AUC \\[4pt]
LODO --- harmonised-reference folds (primary) & Final generalisation across labs, restricted to folds where training and held-out datasets share the same reference type (unipolar-to-Cz vs. bipolar; Section~6.2). Primary FWER-controlled hypothesis tests (H1--H3) are conducted per-dataset within these folds, then macro-averaged (mean across datasets, not pooled raw predictions) to give the test statistic for each metric; the full primary family spans $m=15$ tests (3 hypotheses $\times$ 5 metrics: AUC, BA, Cohen's $\kappa$, per-class F1, MCC), Bonferroni-corrected at $\alpha = 0.05$. Per-dataset values within this primary set are reported descriptively, with FDR (Benjamini-Hochberg) $q$-values as supplementary. & Macro-averaged test AUC (primary) + per-dataset descriptive AUC \\[4pt]
LODO --- full heterogeneous (secondary) & All 6 open-access EDF subsets, regardless of reference type. Reported descriptively only, to characterise the magnitude of the referencing confound; does not feed into the primary FWER-controlled hypothesis family. & Descriptive test AUC only \\
\bottomrule
\end{tabular}
\caption{Validation scheme mapping (revised). The primary hypothesis-testing statistic is now macro-averaged across per-dataset AUCs (never a raw pooled/concatenated-prediction AUC, which risks Simpson's-paradox artefacts), and the primary Bonferroni family spans all hypothesis-metric combinations ($m=15$) rather than metrics alone ($m=5$) or the original invalid per-dataset-vs-aggregate comparison ($m=30$).}
\end{table}

\subsection{PHINN-EEG Generative Model}

We adapt the PHINN rectified flow architecture~[6] for multi-channel EEG. The model learns a velocity field $v_\theta$ mapping Gaussian noise to dream-state EEG, conditioned on a topological conditioning vector $\beta^*$.

\paragraph{Clarification of the conditioning signal (correcting an inconsistency in an earlier draft).} $\beta^*$ denotes the full $\approx 340$-dimensional topological feature vector extracted per epoch in Section~4.2(iii) --- Dynamic Betti Curves, persistence landscapes, persistent entropy, Euler characteristic trajectory, and CPR --- not solely the three Betti curves $\beta_0(t), \beta_1(t), \beta_2(t)$ as the notation might otherwise suggest. This clarification resolves the previous inconsistency between the compact notation here and the explicit statement in Section~5.2 that the spectral-conditioned baseline is dimensionality-matched to the full $\approx 340$-dimensional topological conditioning vector: both the topology-conditioned model and the spectral-conditioned baseline are conditioned on $\approx 340$-dimensional vectors of their respective (topological vs. spectral) feature types, and the comparison in Section~5.2 is between these two full feature-vector representations, not between a full spectral vector and a compact 3-curve topological summary.

\begin{equation}
v_\theta(z_t, t; \beta^*) = v_\theta\big(z_t, t, \text{CrossAttn}(z_t, c^*)\big), \quad \text{where } c^* = \text{FeatureEncoder}(\beta^*),\ \beta^* \in \mathbb{R}^{340}
\end{equation}
(the full topological feature vector, Section~4.2(iii))

\paragraph{Architecture details for $v_\theta$ (pre-registered, correcting a previously under-specified backbone).} $v_\theta$ uses a 1D-convolutional U-Net backbone operating directly on the multi-channel EEG waveform $z_t$ (not a Transformer-only backbone): 4 downsampling stages and 4 upsampling stages, channel widths $64 \to 128 \to 256 \to 512$ (downsampling) mirrored on the upsampling path, kernel size 5, GroupNorm, SiLU activations, and residual connections at each stage. Timestep $t$ is embedded via sinusoidal positional embeddings (128-dimensional) and injected at every U-Net block via FiLM (feature-wise linear modulation). Because the number of EEG channels $K$ varies from approximately 6 to 18 across datasets (Section~4.1), the input layer treats channels as an explicit convolutional axis via a per-channel $1\times1$ projection to a fixed 64-channel latent width before the first downsampling stage, so the U-Net body itself operates on a fixed internal channel dimension regardless of the dataset's native $K$; a separate model instance is trained and evaluated within each dataset's own train/held-out split (in contrast to the classification pipeline's cross-dataset LODO evaluation, which pools multiple datasets during training), so no cross-dataset channel alignment or spatial interpolation is required for the generative model. The conditioning encoder $c^* = \text{FeatureEncoder}(\beta^*)$ processes the topological feature vector $\beta^*$ --- a single $\approx 340$-dimensional aggregate vector (Section~4.6), not a sequence --- so we use a 4-layer MLP (hidden/output dimension 256, GELU activation, LayerNorm and a residual connection after each layer) rather than a Transformer: applying multi-head self-attention to a single token would produce an attention weight of 1.0 by construction and add no representational capacity beyond a feed-forward network, so an explicit MLP is both more honest about what the module does and marginally more efficient. CrossAttn($z_t, c^*$) is inserted at the bottleneck and at each upsampling stage of the U-Net (4 cross-attention insertions total), each with 4 heads and model dimension matched to the corresponding U-Net stage width. Optimiser: AdamW, learning rate $2 \times 10^{-4}$, cosine decay schedule with 1{,}000-step warmup, weight decay $1 \times 10^{-5}$. Batch size: 16. The same architecture, training procedure, and hyperparameters are used for the spectral-conditioned baseline (Section~5.2, item 7) and the unconditional baseline (item 6), with only the conditioning pathway ($\beta^*$ vs. the spectral vector vs. no conditioning) varying, so that architectural capacity is held constant across the three flow-matching comparators.

Training uses:

$L_{\text{train}} = L_{\text{RM}}$ (the rectified flow matching loss~[13]) as the sole training objective for the primary comparison across all conditioning arms (topology-conditioned, spectral-conditioned, unconditional, EEG-GAN, VAE, and diffusion baselines; Section~5.2). We do not additionally include the spectral band-power loss $L_{\text{stat}}$ used in earlier drafts, because $L_{\text{stat}}$ specifically reinforces spectral plausibility and therefore, if shared across arms, would asymmetrically privilege the spectral-conditioned baseline (which targets spectral realism both through its conditioning vector and through a matching loss term) over the topology-conditioned model (whose conditioning target has no analogous loss-level reinforcement, since Vietoris--Rips filtration is non-differentiable and $L_{\text{topo}}$ cannot be backpropagated; see below). Training every arm on $L_{\text{RM}}$ alone ensures that architectural capacity, training procedure, and loss are held constant across arms and that only the conditioning pathway ($\beta^*$ vs. the spectral vector vs. no conditioning) varies, isolating the comparison to conditioning-representation content. As a separate, descriptive-only diagnostic (not part of the primary H4 comparison), we additionally report a variant of the spectral-conditioned baseline trained with $L_{\text{train}} = L_{\text{RM}} + \mu \cdot L_{\text{stat}}$ ($\mu = 0.05$, matching earlier drafts) to characterise, for interpretability, how much of that baseline's spectral coherence is attributable to loss-level reinforcement versus its conditioning vector alone.

\paragraph{Critical implementation note.} Because standard Vietoris--Rips filtration is non-differentiable, topological features are pre-computed offline and used as conditioning inputs to the flow model without backpropagating through the filtration step. $L_{\text{topo}}$ (persistence landscape distance and Euler characteristic matching) is therefore a validation metric and conditioning-fidelity target only --- not a differentiable training loss --- and does not appear in $L_{\text{train}}$ above. A coefficient $\alpha = 0.1$ is retained in the registered protocol as the weight that would apply to a differentiable $L_{\text{topo}}$ term were a differentiable topology layer (e.g., Br\"uel-Gabrielsson et al.) substituted in future work, which would then need to be applied symmetrically (i.e., also adding a matching loss term for the spectral arm, or omitting both) to avoid reintroducing the same asymmetry; under the present non-differentiable implementation, $\alpha$ has no effect on training, and $L_{\text{train}}$ reduces exactly to $L_{\text{RM}}$. Loss coefficients: $\alpha = 0.1$ (reserved for future differentiable extension only, inactive in the present pipeline); $\mu = 0.05$ (used only in the descriptive $L_{\text{stat}}$-augmented diagnostic variant above, not in the primary comparison). Note that these coefficients govern the generative model only; they have no effect on the classification models, which operate on extracted features.

\subsection{Candidate Betti Transition Patterns (Exploratory, Pre-Validation)}

Rather than presenting a formalised atlas at this stage, we outline four candidate qualitative associations between Betti transition patterns and phenomenological dream categories, intended as an exploratory hypothesis space for the empirical validation in Section~4.6 --- not as an established or pre-validated classification scheme. The candidate associations are motivated by Integrated Information Theory~[2] and Global Workspace Theory~[3]: both frameworks predict that richer, more integrated conscious content (e.g., volitionally structured dream narratives) should correspond to network states with sustained, large-scale integration, while fragmented or absent conscious content should correspond to disintegrated, weakly-coupled network states. Persistent $\beta_1$ loops are treated as a candidate geometric correlate of sustained cross-channel integration (a single dominant phase-coupled cycle), and high, unstable $\beta_0$ as a candidate correlate of disintegration (many weakly-coupled components) --- consistent with, but not derived from, IIT/GWT. Table~2 summarises the four candidate patterns at a coarse, qualitative level deliberately, to avoid over-specifying numerical thresholds ahead of any empirical evidence; the precise Betti-curve criteria used to test each pattern (e.g., exact persistence thresholds) will be defined and locked only once the empirical pipeline in Section~4.2 has been run, consistent with the pre-registration logic used elsewhere in this paper. Figure~2 provides a schematic, non-quantitative visualisation. We deliberately avoid presenting this as a finished ``atlas'' until at least the epoch-level validation in Section~4.6 has been completed.

\begin{table}[htbp]
\centering
\small
\begin{tabular}{p{2.4cm}p{4.6cm}p{3.2cm}p{3.4cm}}
\toprule
\textbf{Candidate Pattern} & \textbf{Qualitative Betti Trend} & \textbf{Candidate Dream State} & \textbf{Coarse Report Category} \\
\midrule
Fragmented & $\beta_0$ elevated and variable; $\beta_1$ near zero & No recall / dreamless & No experience reported \\[4pt]
Recurring & $\beta_1$ repeatedly transitions on/off; $\beta_0$ relatively stable & Repetitive / nightmare-like & Experience reported, repetitive theme \\[4pt]
Integrated & $\beta_0$ low and stable; $\beta_1$ sustained; $\beta_2$ near zero & Structured / lucid narrative & Experience reported, structured narrative \\[4pt]
Desynchronised & $\beta_2$ becomes non-zero (void); $\beta_0$ becomes unstable & Bizarre / surreal & Experience reported, bizarreness-rated \\
\bottomrule
\end{tabular}
\caption{Candidate Betti transition patterns (exploratory, pre-validation). Presented at a deliberately coarse, qualitative level; exact numerical criteria are not pre-specified here and will be defined only once the empirical validation in Section~4.6 has been run. This is not a validated classification scheme.}
\end{table}

\begin{figure}[htbp]
\centering
\includegraphics[width=0.85\textwidth]{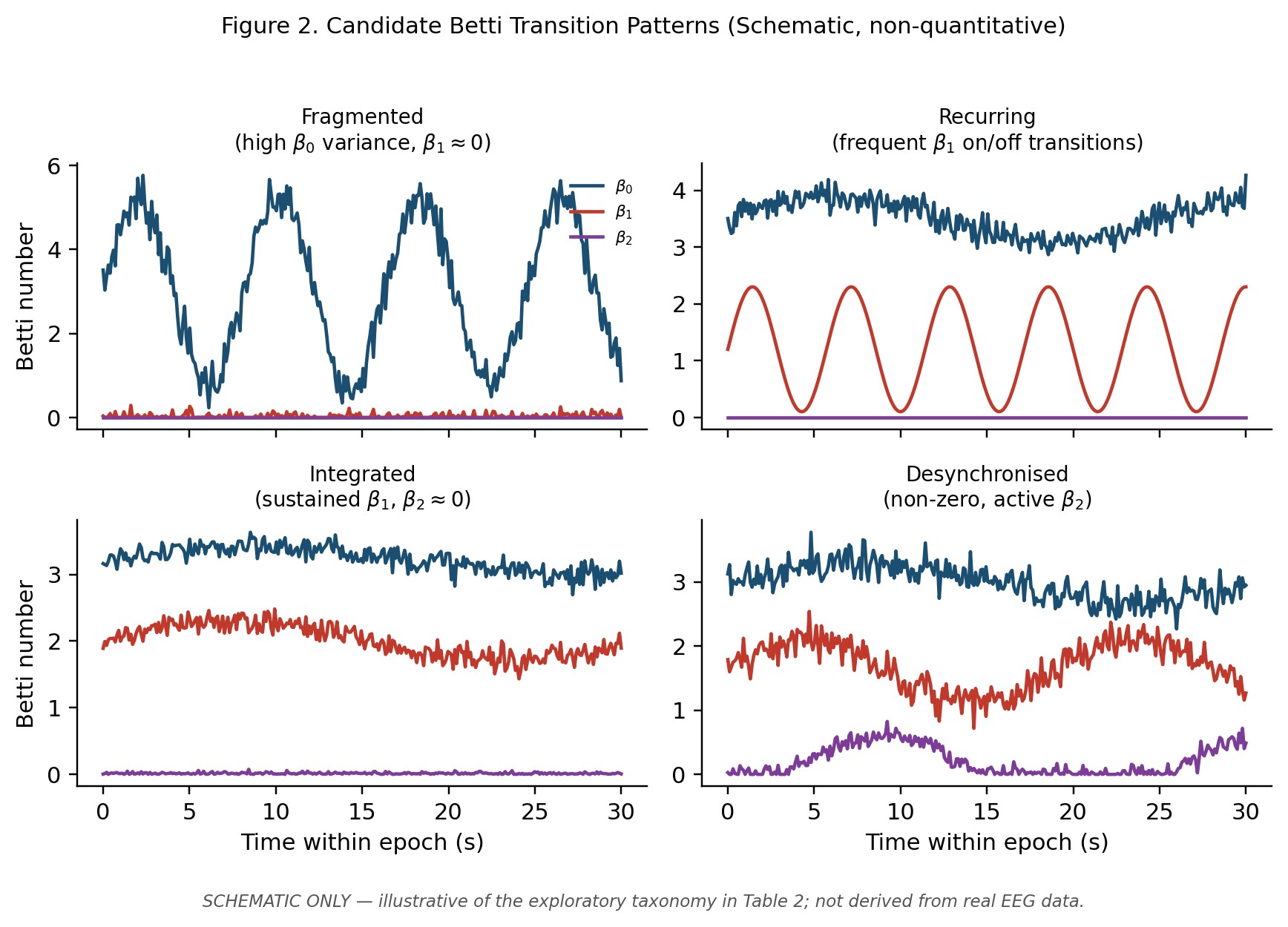}
\caption{Candidate Betti Transition Patterns (Schematic, non-quantitative). All patterns are theoretically motivated, exploratory priors requiring empirical validation (Section~4.6); phenomenological associations are conjectured, not established, and should not be read as a finished atlas.}
\end{figure}

\subsection{Candidate Betti Transition Patterns: Validation Methodology}

We describe the empirical validation procedure for the candidate Betti transition patterns introduced in Section~4.5. Table~2 presents a priori, deliberately coarse pattern definitions requiring post-hoc validation before any more formal scheme is proposed.

\textbf{(i) Epoch-level Betti classification.} The four canonical patterns in Table~2 are defined qualitatively (e.g., ``$\beta_0$ elevated and variable; $\beta_1$ near zero''), so an $L_2$ distance cannot be computed directly against them; we instead operationalise each qualitative pattern as a small set of scalar summary statistics computed from the same Dynamic Betti Curves used elsewhere in the pipeline, and match on those summary statistics rather than on the raw curves. Concretely, for each epoch we compute: mean and variance of $\beta_0(t)$, $\beta_1(t)$, and $\beta_2(t)$ over the epoch; the number of $\beta_1$ on/off transitions (sign changes of the indicator $\mathbf{1}[\beta_1(t)>0]$); the number of $\beta_2$ on/off transitions (sign changes of $\mathbf{1}[\beta_2(t)>0]$); and the fraction of the epoch for which $\beta_1(t)>0$ and, separately, for which $\beta_2(t)>0$. Each of the four qualitative patterns in Table~2 is translated into a corresponding rule over this same summary-statistic space (e.g., ``Fragmented'' = high $\beta_0$ variance with near-zero mean $\beta_1$ and few or no $\beta_1$ transitions; ``Recurring'' = a high $\beta_1$ transition count with stable $\beta_0$ variance; ``Integrated'' = near-zero mean $\beta_2$ and near-zero $\beta_2$-active fraction alongside sustained $\beta_1$; ``Desynchronised'' = a non-zero, non-negligible $\beta_2$-active fraction) prior to seeing any test-set data, with the specific thresholds for ``high'', ``near-zero'', ``few'', ``stable'', and ``non-negligible'' fixed on a held-out calibration split disjoint from the test epochs used for the linguistic association analysis below. Each test epoch's summary-statistic vector is then assigned to the archetype whose region it falls in, or to the nearest archetype centroid by Euclidean distance in this summary-statistic space if the rule regions leave it unclassified; a small set of calibration-epoch assignments will be manually spot-checked against domain-expert visual inspection of the raw Betti curves before the archetype labels are used in any downstream analysis, to catch cases where the summary statistics fail to reflect the intended qualitative pattern. This calibration procedure, and the specific thresholds it produces, will be reported alongside the empirical results. Final archetype-category association testing is performed on held-out test epochs only.

\textbf{(ii) Linguistic feature extraction.} From available DREAM awakening reports: word count (proxy for recall richness), valence and arousal scores using LIWC-22 (licensed; VADER substituted if unavailable), bizarreness rating (where available), and report category label (Experience / No Experience). Bizarreness ratings in particular are expected to be available for only a subset of the 1{,}462 epochs, since not all source studies in the DREAM database collected them; for the mixed-effects and GEE models in (iii) below, we use listwise deletion only for the specific outcome variable under test in a given model (so a model testing bizarreness excludes epochs lacking it, while a model testing word count or valence retains all epochs with that variable present), report the resulting per-model $n$ alongside each result, and additionally report, as a sensitivity analysis, the same models re-fit on multiply-imputed data (multiple imputation by chained equations, 20 imputations, using the other available linguistic and Betti features as predictors) to check whether the complete-case and imputed estimates agree in sign and approximate magnitude; a material disagreement between the two would be reported as a limitation on the affected result rather than resolved by preferring one method silently.

\textbf{(iii) Statistical association, controlling for sleep stage and within-subject correlation.} Sleep stage (REM vs. NREM) is a major confound: dream recall, bizarreness, and word count all differ systematically by stage, and EEG topology itself differs between stages. We therefore do not test raw marginal associations between archetype and dream category. Continuous associations (Betti features such as $\beta_1$ AUC and $\chi$ variance versus linguistic features such as word count and arousal score) are tested via a mixed-effects model with sleep stage as a fixed-effect covariate and participant as a random intercept, rather than marginal Spearman correlation, to account for the repeated-measures structure of the 1{,}462 epochs drawn from the 201 participants represented in the open-access subset (Section~4.1; multiple awakenings per participant violate independence). The categorical archetype-versus-dream-category association is tested using generalized estimating equations (GEE) with an exchangeable working correlation structure clustered on participant ID and sleep stage included as a covariate, rather than a na\"ive chi-squared test, to avoid pseudoreplication and stage confounding. Bonferroni correction is applied across $m=8$ pre-specified tests. We report $n$, model-based effect sizes (regression coefficients or GEE odds ratios), 95\% bootstrap confidence intervals clustered by participant, and FDR-corrected $p$-values as supplementary.

\textbf{(iv) Surrogate data control.} To confirm that extracted Betti curves reflect genuine nonlinear phase-space geometry rather than artefacts of the linear power spectrum, autocorrelation structure, and cross-channel linear coupling (including volume conduction), we generate Multivariate Iterated Amplitude-Adjusted Fourier Transform (MIAAFT) surrogates for each epoch: phase-randomised multi-channel signals that jointly preserve each channel's individual amplitude distribution and power spectrum together with the cross-channel cross-spectral density matrix (and hence zero-lag linear cross-channel correlation, the primary linear signature of volume conduction), while destroying genuine nonlinear structure, including nonlinear cross-channel coupling. We use MIAAFT rather than independent per-channel (univariate) IAAFT specifically because univariate surrogates, applied separately to each channel, preserve only within-channel linear autocorrelation and destroy cross-channel linear correlation entirely; since volume conduction manifests primarily as zero-lag linear cross-channel correlation, univariate surrogates would lack the very confound they are meant to control for, making them unsuitable for this test. We run the full topological feature extraction pipeline (Sections~3.1--4.2) on both the original epochs and their MIAAFT surrogates. If Betti-curve features fail to discriminate real epochs from their surrogates, this indicates the topological signal is attributable to linear structure (within-channel or cross-channel) already captured by PSD/catch22 and volume conduction, undermining the central premise that topology adds information beyond spectral methods and linear coupling. Discrimination between real epochs and surrogates (real-vs-surrogate AUC via Model~1) is reported descriptively alongside the primary hypothesis tests and is not currently a pre-registered gating criterion for H1/H2, but a null result (real-vs-surrogate AUC not exceeding chance) will be treated as evidence against the topological hypothesis and reported as such.

\textbf{(v) Channel-perturbation control.} To empirically distinguish genuine multi-channel cross-channel integration from single-channel oscillatory dominance (the concern raised in Section~6.1), we recompute joint-embedding topological features under two perturbations, applied independently to each channel in turn while holding the remaining channels intact: (a) IAAFT phase-randomisation of the single target channel only, and (b) ablation (zeroing) of the single target channel. If a persistent $\beta_1$ loop in the joint embedding survives single-channel perturbation or ablation, this is evidence that the loop reflects genuine multi-channel coupling rather than a single dominant oscillator being carried through concatenation; if the loop collapses under perturbation or ablation of one particular channel, this indicates single-channel dominance by that channel. Results are reported per-channel as a supplementary robustness analysis alongside the primary hypothesis tests, and are the analytical basis for any claims of large-scale, multi-channel integration made in Section~6.1.

\textbf{(vi) Current status.} Validation of the candidate patterns is a planned empirical deliverable following implementation of the full PHINN-EEG pipeline. The patterns in Table~2 are theoretically motivated, exploratory priors --- not yet empirically confirmed on the DREAM database, and not to be treated as a validated atlas prior to that confirmation.

% ======================================================
\section{Experiments and Projected Results}

\subsection{Datasets}

Primary analysis uses the 1{,}462 labelled awakening epochs from the DREAM database's six open-access EDF subsets~[1]. Inclusion flow: 3{,}191 total awakenings in the DREAM registry $\to$ 2{,}812 with complete sleep-stage labels $\to$ 1{,}462 with publicly available raw EDF files (open-access subset; used for model training and evaluation). Exclusions at each stage will be reported with a CONSORT-style flow diagram. We apply leave-one-dataset-out (LODO) cross-validation across the 6 open-access EDF subsets to test generalisation across laboratory conditions, electrode configurations, and demographic profiles; as detailed in Section~6.2, the primary hypothesis-testing LODO is restricted to folds sharing the same reference type (unipolar-to-Cz vs. bipolar), with the full heterogeneous-reference LODO reported as a secondary, descriptive analysis. Full LODO across all 20 DREAM contributing studies will require institutional data access agreements not available at Stage~1. The Sleep-EDF Expanded database (197 whole-night recordings, PhysioNet) provides independent validation for the REM-detection component.

\subsection{Baselines}

\paragraph{Classification baselines:} (1) Wong-PSD: PSD-only logistic regression (AUC $=0.586$/NREM, $0.700$/REM). (2) Wong-catch22: catch22 broadband features (AUC $=0.700$/REM). (3) Moctezuma-CSP: CSP + KNN (AUROC $>0.85$). (4) SleepEEGNet: single-channel CNN, 88.7\% REM recall. (5) Model~2 / Model~2b (XGBoost on raw EEG features, no topology) --- primary and dimensionality-matched ablations for H3 (Section~4.3).

\paragraph{Generative baselines:} (6) Unconditional rectified flow (no conditioning). (7) Spectral-conditioned rectified flow: identical architecture to the PHINN-EEG generative model (same rectified flow backbone and $L_{\text{RM}}$-only training objective, Section~4.4) but with the topological input $\beta^*$ replaced, at the FeatureEncoder's input, by a high-resolution spectral feature vector --- full-resolution multi-taper spectral bins plus the full cross-channel coherence matrix across all channel pairs --- constructed to match the $\approx 340$-dimensional feature budget of $\beta^*$, rather than the coarse 6-band power vector used in earlier drafts. This spectral vector is passed through the same FeatureEncoder MLP (4 layers, hidden/output dimension 256; Section~4.4) used for the topological conditioning pathway, yielding the identically-dimensioned 256-dimensional conditioning signal $c^*$ that feeds the downstream cross-attention layers unchanged. This ensures the comparison isolates the representational content of topological versus spectral conditioning rather than confounding it with conditioning-vector dimensionality or with any change to the cross-attention architecture. Architecture, training, and data are held constant; only the conditioning representation and its matched dimensionality vary. (8) EEG-GAN. (9) VAE. (10) Latent diffusion baseline: a representative latent-diffusion model in the style of Aristimunha et al.~[12], trained from scratch on the same per-dataset LODO splits used throughout this study (rather than relying on an unavailable pre-trained checkpoint), included because diffusion-based approaches represent the current state of the art in neural signal synthesis and their omission would leave the generative comparison incomplete. Generative evaluation uses Fr\'echet EEG Distance (FED) and spectral coherence.

\subsection{Evaluation Metrics}

\paragraph{Classification:} AUC, balanced accuracy (BA), Cohen's $\kappa$, per-class F1, Matthews Correlation Coefficient (MCC). For each metric, we first compute the value independently within each harmonised-reference held-out dataset (Section~6.2), then macro-average (unweighted mean) across datasets to obtain the primary test statistic; raw predictions are never pooled across datasets before scoring, to avoid Simpson's-paradox-style artefacts from between-dataset base-rate differences (see Section~3.3). Each macro-averaged metric is compared against the corresponding Wong baseline where available. The primary FWER-controlled family spans all three primary hypotheses (H1, H2, H3) across all five metrics --- $m=15$ tests in total --- Bonferroni-corrected at $\alpha = 0.05$ (not $m=5$, correcting an earlier undercount that considered only the metrics and not the hypotheses). Per-dataset values within the harmonised-reference set are additionally reported descriptively to demonstrate robustness to inter-laboratory heterogeneity, with FDR-corrected (Benjamini--Hochberg) $q$-values as supplementary. Hypothesis H4 (generative quality) is evaluated separately via the bootstrap procedure described below, since its metrics (SCR, FED) are not computed per-dataset in the same way. The full heterogeneous LODO across all 6 subsets (mixing reference types) is reported separately, purely descriptively, to quantify the referencing confound itself, and does not contribute to the primary FWER-controlled family. If full 20-dataset access is obtained in a later stage, the same macro-averaging, harmonised-reference-primary logic will be applied at that scale.

\paragraph{Generative significance testing (H4).} Differences in FED and SCR between the topology-conditioned model and each baseline (unconditional flow, EEG-GAN, VAE, the dimensionality-matched spectral-conditioned flow, and the diffusion baseline) are assessed via a cluster (participant-level) bootstrap rather than resampling individual epochs directly: because the held-out real epochs are drawn from multiple epochs per participant (1{,}462 epochs from 201 participants; Section~4.1), treating epochs as independent resampling units would violate the independence assumption and artificially narrow the confidence intervals. We instead resample, with replacement, the set of participants contributing to the held-out real-epoch pool 2{,}000 times; for each resample, all epochs belonging to a selected participant are included (a participant selected twice contributes their epochs twice), the corresponding generated epochs are paired to the same resampled participant set where the generative model's held-out conditioning draws from real per-participant epochs, each metric is recomputed on the resulting pooled sample, and we report 95\% bootstrap confidence intervals for the pairwise differences computed from the resulting distribution across the 2{,}000 participant-level resamples. To properly apply the stated Bonferroni correction for $m=5$ pairwise baseline comparisons (unconditional, spectral-conditioned, EEG-GAN, VAE, and diffusion) at a family-wise alpha of 0.05, a difference is treated as significant if its 99\% bootstrap CI (corresponding to the Bonferroni-adjusted per-comparison alpha of 0.01) excludes zero, for each of FED and SCR separately.

\paragraph{Topological fidelity:} Betti-RMSE, Persistence Landscape Wasserstein distance, Transition Accuracy (TA), and CPR. Betti-RMSE is computed by re-extracting Dynamic Betti Curves from each generated epoch using the identical pipeline of Section~4.2, and taking the root-mean-square distance between this re-extracted topological feature vector and the input conditioning vector $\beta^*$ used to generate that epoch; because the flow-matching and spectral loss terms used to train the topology-conditioned model (Section~5.2) do not include a topological term (the extraction step is non-differentiable), this held-out re-extraction is the primary direct check of whether the model has in fact learned to synthesise the conditioned topology, rather than an assumption.

\paragraph{Generative quality:} FED, spectral coherence, cross-correlation with held-out real epochs, and Scenario Coverage Rate (SCR).

\subsection{Projected Results}

\textbf{Important:} All performance figures below are explicitly prospective projections grounded in published TDA-EEG benchmarks --- not experimental results. All PHINN-EEG performance claims are labelled as projections until empirical validation on the DREAM database is complete.

\begin{figure}[htbp]
\centering
\includegraphics[width=0.85\textwidth]{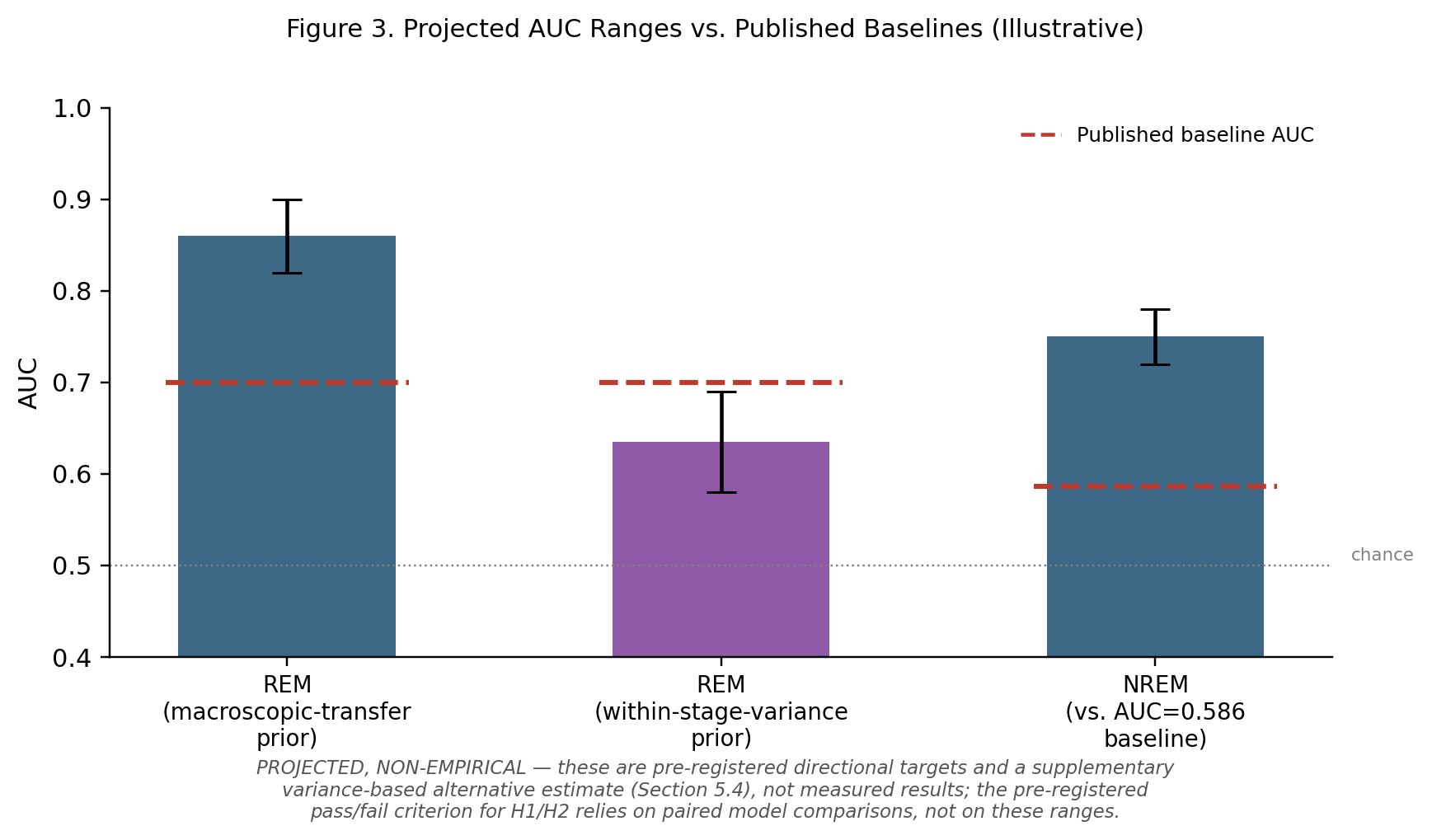}
\caption{Projected AUC ranges vs. published baselines (illustrative; see caption in figure and Section~5.4 text for non-binding status).}
\end{figure}

\paragraph{Classification (REM, primary endpoint).} We project AUC $= 0.82$--$0.90$ for the topology ensemble versus AUC $= 0.700$ (Wong catch22), tested using the macro-averaged procedure defined in Section~5.3. We present this range as an illustrative, weakly-grounded prior rather than a rigorously derived transfer estimate: the underlying evidence comes from topological-feature performance in sleep staging (Baronetzky~[5]) and meditative-vs-resting EEG classification (Gupta et al.~[4]) --- both sources are verified (Section~5.4), but both are macroscopic state-shift classification tasks, not within-stage rare-event detection, and neither directly measures dream-mentation discrimination. We do not claim a sound theoretical or empirical basis for transferring specific effect sizes from these disparate tasks to dream detection; the 0.82--0.90 range should be read as a working target for the pre-registration exercise, not a calibrated prediction. Should~[4] fail to be verified prior to publication, the projection would rest on~[5] and Moctezuma et al.~[7] alone. Because sleep staging and meditative EEG involve substantially higher signal-to-noise ratios than within-stage dream mentation detection, actual topological effect sizes in dream detection may be substantially smaller than either source suggests.

Because the 0.82--0.90 range is only a weakly-grounded prior (above), we do not use a fixed absolute AUC threshold to decide whether the topological hypothesis is supported. Instead, the pre-specified gating criterion for H1 is defined relative to the two comparisons established directly on this task in Section~3.3/5.3: statistically significant superiority of Model~1 over Model~2 (the PSD/catch22 ablation, evaluated on the identical per-dataset LODO folds; the primary paired test defined in Section~3.3) constitutes support for H1, and statistically significant superiority of PCA-reduced Model~1 over native Model~2 (the dimensionality-controlled comparison; H3, Section~5.3) constitutes evidence that this improvement is attributable to topological content specifically rather than to feature-space capacity alone. A result that clears Model~2 (full-dimension Model~1) but not the PCA-reduced H3 comparison would indicate genuine but non-topology-specific improvement (attributable to feature-space capacity rather than topological content); a result that clears neither is a null result. Clearance of the externally published Wong catch22 benchmark (AUC $=0.700$) is reported alongside as descriptive corroboration, consistent with Section~3.3, but --- since its performance was established under a different evaluation protocol --- is not itself part of the pass/fail criterion. The 0.82--0.90 range remains useful for characterising how a given outcome compares against the (weakly-grounded) prior, but is not used as the pass/fail criterion for the pre-registered hypothesis.

As a supplementary, more directly relevant reference point, we additionally derive a within-stage-variance-based plausible range: given typical REM dream-recall base rates of 50--80\% upon awakening (implying a near-balanced binary discrimination target) and assuming the topological feature set captures a small-to-moderate fraction of the between-condition variance typical of subtle within-stage EEG contrasts (Cohen's $d \approx 0.3$--$0.5$, consistent with within-stage neural correlates reported in the dream-EEG literature rather than the macroscopic-state-shift literature), a rough analytic AUC range of 0.58--0.69 (via $\text{AUC} = \Phi(d/\sqrt{2})$ for the assumed $d$ range) is obtained; this is substantially lower than, and in tension with, the 0.82--0.90 macroscopic-transfer prior, and is reported alongside it precisely to make visible how sensitive the projected range is to the choice of analogy. We do not resolve this tension by selecting one figure over the other; both are retained as descriptive, non-binding context, and the pre-registered pass/fail criterion for H1 relies solely on the paired model comparisons described above, not on either projected range.

\paragraph{Classification (NREM).} We project AUC $= 0.72$--$0.78$ versus AUC $= 0.586$ (Wong PSD). NREM dreams are more fragmented and the signal is expected to be noisier.

\paragraph{Ablation (H3).} PCA-reduced Model~1 (topological features, reduced to Model~2's native $\approx 220$-dimensional budget via per-fold training-only PCA) versus native Model~2 (raw EEG features, $\approx 220$ features) --- same feature-set size without padding either model, as the primary dimensionality-controlled comparison; Model~2b (Model~2 padded to $\approx 340$ features) and native-dimension Model~1 versus Model~2 are reported descriptively for continuity with the original protocol. Projected AUC reduction of 6--12 points for Model~2 relative to PCA-reduced Model~1, confirming a topology contribution independent of both model capacity and feature-set size.

\paragraph{Generation.} Topology-conditioned model projected to achieve SCR $= 60$--$70\%$ (vs 3--5\% random baseline), lower FED than the unconditional flow, spectral-conditioned flow, EEG-GAN, VAE, and diffusion baselines (Section~5.2, items 6--10, tested via the bootstrap procedure in Section~5.3), and spectral coherence within 5\% of real REM epochs across all six bands. Because a pre-trained Aristimunha et al.~[12] checkpoint is not available for the DREAM open-access subsets, and that work targets general sleep-stage EEG generation on a different dataset and evaluation protocol, our diffusion baseline (item 10) is a representative latent-diffusion architecture trained from scratch on the same per-dataset LODO splits used throughout this study, rather than a reproduction of the Aristimunha et al. checkpoint itself; it is included as a like-for-like diffusion comparator evaluated under our own protocol, not as a direct replication of their reported numbers. We continue to note Aristimunha et al.~[12] in Section~2.3 as related generative work.

The dimensionality-matched spectral-conditioned baseline is expected to achieve higher spectral coherence but lower topological fidelity (i.e., higher Betti-RMSE) as measured by Betti-RMSE against the conditioning target; because this Betti-RMSE comparison is not fully independent of the conditioning signal each model receives (Section~5.2), we additionally evaluate both generative models using two downstream, conditioning-blind checks rather than one, to avoid a structural bias toward whichever generator shares its feature representation with the evaluator. First, real and generated epochs (from both the topology-conditioned and spectral-conditioned models) are pooled and classified by Model~1 (topological features only, trained only on real data) without access to which model generated which synthetic epoch or to either model's conditioning vector. Second, the same pooled real/synthetic epochs are independently classified by Model~2 (raw/spectral EEG features only, trained only on real data), under the same blinding. Because Model~1 operates exclusively on topological features, it is structurally more likely to judge the topology-conditioned generator's output as realistic simply because both share a feature representation, independent of whether the underlying raw EEG waveform is realistic; the Model-2 check uses a feature representation disjoint from the topology-conditioned model's conditioning signal and is therefore not subject to this bias, and serves as the primary arbiter of whether the topology-conditioned model produces holistically realistic raw signals rather than merely overfitting the topological feature space. A generative model whose synthetic epochs are harder for the relevant classifier to distinguish from real epochs, and whose synthetic epochs receive feature values closer to the real distribution, is judged to have produced more realistic output by that classifier's criterion; we treat the two checks as complementary rather than collapsing them into a single score, and report both.

% ======================================================
\section{Discussion}

\subsection{Why Topology Outperforms Spectral Methods}

Power spectral density is blind to phase relationships. Two EEG epochs can have identical band power profiles yet fundamentally different phase coupling structure --- coherent alpha oscillations (a stable limit cycle, yielding persistent $\beta_1 = 1$) versus fragmented, incoherent alpha activity (a diffuse cloud, yielding high $\beta_0$ with no loops). This difference is invisible to PSD but immediately captured by persistent homology. The catch22 feature set captures nonlinear statistical moments but not global topological connectivity.

We note, consistent with the caveat in Section~3.1, that a persistent $\beta_1$ loop in the concatenated joint embedding is not on its own diagnostic of large-scale, multi-channel integration. This can happen in at least two distinct ways: (a) a strong single-channel limit cycle (e.g., localized occipital alpha) can produce a $\beta_1$ loop in the joint phase space even if other channels remain desynchronised, and (b) more generally, concatenation of independent, uncoupled per-channel oscillators --- with no true cross-channel dynamical relationship at all --- can itself produce apparent loop structure in the joint point cloud purely as a geometric artefact of stacking multiple periodic signals into a higher-dimensional space, independent of any single channel dominating. Neither case constitutes genuine multi-channel integration, and our use of ``large-scale fronto-parietal integration'' language should be read as flagging a candidate interpretation under test, not an established mechanistic reading of $\beta_1 > 0$. Distinguishing genuine cross-channel integration from both failure modes --- single-channel dominance and independent-oscillator concatenation artefacts --- requires the channel-perturbation control described in Section~4.6(v) (which addresses single-channel dominance directly) together with the MIAAFT surrogate control (Section~4.6(iv), which independently phase-randomises each channel and so would also disrupt spurious loops arising purely from concatenated independent oscillators); claims of large-scale integration in this paper should be read with both qualifications throughout.

\subsection{Sparse Channels and Global Topological Claims}

A theoretical tension exists between our claim that PHINN-EEG detects large-scale fronto-parietal integration and the practical requirement to harmonise across heterogeneous DREAM datasets using a minimal common electrode set.

Where raw unipolar recordings are available, all channels are re-referenced to a common Cz (or linked-mastoid) scheme prior to further processing, yielding a consistent common-referenced signal set for that subset (Section~4.1). This re-referencing is applied only where unipolar data are available; it is not attempted where a dataset provides only pre-calculated bipolar derivations. Several of the DREAM open-access subsets (following Wong et al.) provide exactly such pre-calculated bipolar derivations --- for example Fpz-Cz and C4-M1 --- which are themselves already differential signals, not unipolar channels; recovery of their unipolar components from the differential signal alone is mathematically underdetermined, so re-referencing to Cz is not attempted for these subsets, consistent with Section~4.1. For the open-access DREAM subsets that provide only pre-calculated bipolar channels, the topological pipeline is instead adapted to operate directly on the bipolar signals without attempting re-conversion to a common unipolar reference (see Sections~4.1 and~6.5). Because this means the delay-embedded point cloud is constructed from mathematically different signal types (common-referenced unipolar potentials in some datasets, differential bipolar signals in others), the resulting phase-space geometries are not guaranteed to be directly comparable across datasets. This is a genuine limitation of LODO validity across heterogeneous montages: LODO folds that pair a unipolar-referenced training set with a bipolar-referenced held-out dataset (or vice versa) test generalisation across both dream content and a change of signal representation simultaneously, which may inflate apparent generalisation error unrelated to the topological hypothesis itself. We report LODO results per-dataset (Section~5.3) specifically so that this confound can be inspected post hoc, and we flag datasets by reference type in the released data splits.

\paragraph{Primary evaluation restricted to harmonised-reference folds.} Per the concern above, the primary LODO evaluation used for all FWER-controlled hypothesis tests (H1--H3, Section~5.3) is restricted to folds where both the training and held-out datasets share the same reference type (unipolar-to-Cz vs. bipolar); folds that pair a unipolar-referenced training set with a bipolar-referenced held-out dataset (or vice versa) are excluded from the primary analysis, since such folds confound the topological hypothesis with a change of signal representation. The unipolar-to-Cz subsets number four of the six open-access EDF subsets, and the bipolar-derivation subsets number two; the primary LODO is therefore run separately within each reference-type group (each dataset held out in turn against the other datasets of the same reference type), and the macro-averaged AUC used for hypothesis testing (Section~5.3) is computed as the unweighted mean of the resulting per-dataset AUCs within these harmonised-reference folds --- never as a single AUC computed from raw predictions pooled across folds, which would risk Simpson's-paradox-style artefacts from between-dataset base-rate differences. As a secondary, purely descriptive analysis, we additionally report the full heterogeneous LODO across all 6 subsets regardless of reference type, to characterise how much apparent generalisation error is attributable to the referencing confound itself; this full heterogeneous analysis does not feed into the primary FWER-controlled hypothesis family. Average re-referencing is not used, as it introduces systematic confounds across datasets with heterogeneous electrode densities (2--256 channels) where the average reference is not mathematically equivalent.

Our projection of AUC $= 0.82$--$0.90$, using the full joint point cloud constructed from all available channels per dataset ($K$ ranging from approximately 6 to 18 across the primary open-access subsets, per Section~4.1--4.2), is grounded in Moctezuma's AUROC $>0.85$ using 8--10 channels with spatial pattern features~[7]; topological features may achieve comparable performance at these channel counts by capturing different geometric information than spatial-pattern methods.

The CPR metric (defined formally in Section~4.2(iii)) quantifies robustness of topological features to point-cloud perturbation via a bootstrap noise threshold. When CPR is examined in the specific context of channel reduction discussed in this section, it is computed on the point cloud resulting from the reduced-channel joint embedding, using the same bootstrap perturbation-threshold definition as in Section~4.2(iii); channel reduction is therefore treated as one specific instance of point-cloud perturbation for CPR purposes, not a separate metric.

\subsection{BCI Relevance and Real-Time Feasibility}

SleepEEGNet demonstrates that REM detection from a single EEG channel is solved (88.7\% recall)~[8]. PHINN-EEG is intended to extend the state of the art for BCI deployment through minimal-channel joint embeddings compatible with consumer-grade forehead EEG devices; as noted in Section~5.4, this remains a projection pending empirical validation, not an established result. This BCI-deployment motivation is in tension with a limitation already noted in Section~6.2: because the topological pipeline operates on mathematically different signal types (unipolar-referenced vs. bipolar-differential), the resulting phase-space geometries are not directly comparable across reference schemes, and our primary evaluation is consequently restricted to harmonised-reference folds (training and held-out data sharing the same reference type). This restriction is a genuine limitation on the claim of generalising across the full heterogeneous DREAM database, and it matters specifically for BCI deployment, where consumer-grade forehead devices, clinical unipolar montages, and bipolar research montages all vary in reference scheme and a deployed model may see a reference type it was not trained on. We identify three concrete steps toward closing this gap, none of which we have yet implemented or validated: (i) a reference-invariant preprocessing step --- for example, converting all subsets to a common average or a small set of anchor bipolar derivations shared across montages, so that downstream topological features are computed from a common reference frame rather than the dataset's native one; (ii) explicitly training and reporting a cross-reference-scheme LODO condition (already recorded descriptively as the ``full heterogeneous LODO'' in Section~5.3, but not currently used to support any primary hypothesis) as a primary, not merely descriptive, generalisation metric in future work; and (iii) reference-scheme domain adaptation (e.g., adversarial or feature-alignment methods analogous to those used for inter-subject BCI transfer~[23]) applied specifically to the joint delay-embedding features, to test whether the reference-scheme gap can be closed algorithmically rather than only by restricting evaluation to harmonised folds. Until one or more of these is implemented and validated, claims of cross-laboratory or cross-device generalisation should be read as applying only within a shared reference scheme, not across the full heterogeneity of the DREAM database or of real-world BCI hardware.

\paragraph{Real-time feasibility.} Betti curve extraction on a Jetson AGX Orin achieves $<98$~ms latency for single-channel embeddings (per PHINN~[6]). For an 8-channel joint filtration (illustrative $K$ within the 6--18 channel range; Section~4.1) at the default $d=7$, per sub-window ($\approx 500$-point clouds in $\mathbb{R}^{56}$; Section~4.2), the landmark sub-sampling protocol and the corrected 10th-percentile filtration scale (Section~3.2) are estimated to keep single-sub-window latency under 300~ms; the higher candidate dimensions in the sensitivity sweep ($d=10,15$) are expected to increase per-distance computation cost roughly linearly in $d$ without changing the underlying $O(n^2)/O(n^3)/O(n^4)$ simplex-count scaling (Section~4.2(ii)), so this latency estimate is not expected to change qualitatively across the sweep, though it will be measured empirically for each candidate $d$ and channel count $K$. However, because filtration is now performed independently on each of the $\approx 26$ overlapping sub-windows per 30-second epoch (Section~4.2), the total feature-extraction cost for a full epoch is on the order of several seconds on target hardware, not real-time within the epoch's own duration. Sub-window-level latency ($<300$~ms) is compatible with streaming, online Betti-curve updates at the 1~Hz stride rate, but full-epoch real-time feasibility as originally framed is not supported by the current pipeline. Empirical timing on target hardware will be reported in the experimental follow-up, and real-time feasibility claims are restricted to sub-window-level, streaming inference only, pending those results.

\subsection{Ethical Considerations}

Dream monitoring raises significant ethical considerations around cognitive privacy. We align with emerging neurorights frameworks~[19] and recommend: (i) all deployed systems operate with explicit, revocable informed consent; (ii) dream topology data be treated as sensitive health information under GDPR; (iii) closed-loop stimulation protocols be subject to institutional ethics review.

\subsection{Limitations and Future Work}

\textbf{First, electrode harmonisation:} For the 1{,}462 open-access EDF subsets, we first verify whether raw unipolar data are available. Where only bipolar derivations (Fpz-Cz, C4-M1) are provided, the topological pipeline operates directly on bipolar signals as differential measures, rather than attempting the mathematically impossible recovery of absolute unipolar potentials from differential signals. See Section~6.2 for the full protocol and for discussion of the resulting cross-dataset comparability limitation.

\textbf{Second,} dream reports are subjective and linguistically variable; the candidate Betti transition patterns (Section~4.5) require empirical validation (Section~4.6) and expansion to datasets with standardised phenomenological reporting before any more formal atlas is proposed.

\textbf{Third,} higher homology dimensions ($H_3$ and above) may capture cross-frequency phase-amplitude coupling, but are computationally prohibitive with current VR implementations.

\textbf{Fourth,} delay embedding creates temporally autocorrelated point cloud vectors: successive delay vectors are formed by advancing the raw signal by a single sample, so their underlying sample spans overlap substantially in time even though a lag $\tau > 1$ (set to the pilot-estimated first zero-crossing of the autocorrelation function, Section~3.1) means they do not share identical sample values. This temporal proximity, combined with the signal's own autocorrelation structure, induces correlation between nearby delay vectors; standard persistent homology assumes no such temporal structure, and this autocorrelation can produce spurious topological features. Our sliding sub-window grid (Section~4.2) itself overlaps by 80\% between consecutive sub-windows, so this autocorrelation concern applies both within a sub-window's point cloud and, to a lesser extent, between the Betti-curve estimates of adjacent overlapping sub-windows. We address the within-window concern via the MIAAFT surrogate-data control (Section~4.6(iv)) and by reporting CPR across bootstrap resamples; we additionally address the sub-window-overlap component directly via the pre-specified overlap sensitivity analysis described in Section~4.2(i) --- spanning 0\% (non-overlapping), 50\%, 65\%, and 80\% overlap --- which empirically tests whether classification performance depends on the degree of induced autocorrelation rather than assuming the concern away; the 0\% (non-overlapping) setting in particular provides a baseline structurally immune to this specific confound. A fully formal statistical correction for autocorrelation-induced topological artefacts in persistent homology remains an open problem in the TDA literature more broadly and is out of scope for this pipeline; the overlap sensitivity analysis is intended as an empirical, rather than theoretical, mitigation.

\textbf{Fifth,} DREAM awakening reports are collected after awakening, introducing temporal misalignment: the classified EEG window captures awakening dynamics as well as residual dream-state activity. We adopt the same 30-second pre-awakening window as Wong et al.~[1] for benchmark comparability, but future work with intra-sleep awakening protocols would better isolate the dream-content EEG signal.

To partially disambiguate dream-state topology from arousal-onset topology within this fixed window, we will additionally report Betti curves separately for the early (0--15~s) and late (15--30~s) halves of each pre-awakening epoch as a descriptive, non-primary analysis; a topological signature concentrated in the final seconds immediately preceding awakening (rather than sustained across the full window) would be consistent with an arousal-driven rather than dream-content-driven effect, and will be flagged as such in reporting even though it does not resolve the underlying confound.

\textbf{Sixth,} aggregating 20 heterogeneous studies introduces uncontrolled confounds (amplifier type, sampling rate, reference scheme); future work should investigate EEG harmonisation methods (e.g., ComBat) applied prior to topological feature extraction.

\textbf{Seventh, adaptive per-window filtration scale:} Setting $\varepsilon^*(t)$ to the 10th percentile of pairwise distances within each individual sub-window (Section~3.2) normalises the Rips filtration relative to that sub-window's own internal variance. This keeps the complex sparse across heterogeneous point-cloud densities, but it also means the resulting Betti curves are scale-free with respect to absolute signal amplitude and power: a high-amplitude, structured oscillation and a low-amplitude noise process could in principle yield similar Betti curves if their internal geometries have similar relative structure. This is a deliberate design choice to isolate shape from energy (Section~1), but it also means PHINN-EEG topological features should be interpreted as complementary to, not a full replacement for, absolute-power features when distinguishing sleep stages with very different absolute signal energies (e.g., high-amplitude slow-wave NREM versus low-amplitude desynchronised REM).

To further guard against the adaptive scale artificially manufacturing structure from unstructured input, we run two complementary controls. First, we run the full pipeline on simulated scale-free ($1/f$) noise epochs matched in length and channel count to real EEG; if this noise-only control yields Betti curves with sustained $\beta_1 > 0$ comparable to real dream-state epochs, this would indicate the adaptive threshold generates spurious loop structure rather than detecting genuine dream-state topology, and would be reported as a limitation of the adaptive-scale design regardless of classification performance on real data. Second, and directly addressing the reviewer's concern that adaptive normalisation could equate a high-amplitude structured signal with a low-amplitude noise process of similar relative geometry, we run the pre-specified fixed-scale (per-fold-calibrated) parallel analysis described in Section~4.2(i) on the real classification task itself; comparable REM AUC between the adaptive-scale and fixed-scale pipelines would support the interpretation that classification performance reflects genuine topological structure rather than an artefact of local-density normalisation, whereas a large discrepancy would indicate the adaptive scale is inflating apparent performance. We retain PSD and catch22 features in Model~2/2b and report per-stage performance (Section~5.3) so that any stage-dependent artefacts introduced by this normalisation can be identified empirically rather than assumed away.

\textbf{Eighth,} as detailed in Section~3.1 and Section~4.1, we do not apply a surface Laplacian (or other dense-montage spatial control) to disambiguate volume-conduction-driven zero-lag channel mixing from genuine lagged dynamical coupling, because the spherical-spline method requires a dense unipolar montage ($\geq 32$ channels) that none of the primary open-access subsets provide (they range from approximately 6 to 18 channels). This concern is therefore unresolved for all primary subsets, unipolar and bipolar alike, not only the bipolar ones; the MIAAFT surrogate control (Section~4.6(iv)) and the channel-perturbation control (Section~4.6(v)) remain the only available (partial, indirect) evidence across the full dataset. Identifying or developing a volume-conduction control suited to sparse montages --- e.g., a low-order local (nearest-neighbour) Laplacian requiring only 4--6 spatially adjacent electrodes, or a bespoke sparse-array source-separation method --- is identified as a priority methodological extension for the empirical follow-up, to be applied only where a given subset's electrode layout can support it.

% ======================================================
\section{Conclusion}

We have introduced PHINN-EEG, the first topological time-series framework for EEG-based dream mentation classification and synthesis. By replacing power spectral density features with Dynamic Betti Curves derived from multi-channel Takens delay embeddings, we access a feature space characterising the geometric architecture of neural activity rather than its energy. This shift is theoretically motivated and analytically supported primarily by the verified benchmark of Baronetzky~[5] (a topological-versus-statistical accuracy advantage in sleep-stage classification), with Gupta et al.~[4] cited as a corroborating secondary source from a related EEG domain (both references verified under corrected bibliographic details; Sections~1, 5.4).

The DREAM database~[1] establishes a clear benchmark at AUC $= 0.70$ for REM dream detection. PHINN-EEG targets AUC $= 0.82$--$0.90$, with a pre-specified interpretation plan for results below this range. Realising this target requires empirical execution of the proposed pipeline, which constitutes our immediate next step. The candidate Betti transition patterns introduced here provide a starting, exploratory geometric vocabulary for neural dream phenomenology; formalising them into a validated atlas is left for future work, pending the empirical validation described in Section~4.6. All code, pre-trained weights, and topological feature extraction utilities will be released upon publication.

This work represents a paradigm shift from asking ``how much brain activity?'' to ``what shape is the brain's activity?'' in the context of conscious dreaming.

% ======================================================

\vspace{1em}
\noindent\textbf{Author Contributions.} R.T.: Conceptualisation, methodology. E.Y.: theoretical framework, writing. J.B.: review and supervision.

\vspace{0.5em}
\noindent\textbf{Competing Interests.} The authors declare no competing interests. No external funding was received.

\end{document}